%% file: main.tex
\documentclass[%
 aip,
 amsmath,amssymb,
 reprint,%
]{revtex4-1}
\input{packages.tex}
\input{macro.tex}

\usepackage{graphicx}%
\usepackage{dcolumn}%
\usepackage{bm}%

\usepackage[utf8]{inputenc}
\usepackage[T1]{fontenc}
\usepackage{mathptmx}
\usepackage{etoolbox}
\DeclareMathAlphabet{\mathcal}{OMS}{cmsy}{m}{n}
\newcommand{\name}{\texttt{SP\textsuperscript{2}RINT}\xspace}
\makeatletter
\def\@email#1#2{%
 \endgroup
 \patchcmd{\titleblock@produce}
  {\frontmatter@RRAPformat}
  {\frontmatter@RRAPformat{\produce@RRAP{*#1\href{mailto:#2}{#2}}}\frontmatter@RRAPformat}
  {}{}
}%
\makeatother
\begin{document}

\title{SP$^2$RINT: \underline{S}patially-Decoupled \underline{P}hysics-Inspired \underline{P}rog\underline{R}essive \underline{IN}verse Op\underline{T}imization for Scalable, PDE-Constrained Meta-Optical Neural Network Training}

\author{Pingchuan Ma}
\affiliation{ 
School of Electrical, Computer and Energy Engineering, Arizona State University, Tempe, AZ 85281, USA%
}

\author{Ziang Yin}
\affiliation{ 
School of Electrical, Computer and Energy Engineering, Arizona State University, Tempe, AZ 85281, USA%
}
\author{Qi Jing}
\affiliation{ 
School of Electrical, Computer and Energy Engineering, Arizona State University, Tempe, AZ 85281, USA%
}

\author{Zhengqi Gao}
\affiliation{ 
Department of Electrical Engineering and Computer Science, Massachusetts Institute of Technology, Boston, MA 02139, USA
}

\author{Nicholas Gangi}
\affiliation{ 
Department of Electrical, Computer, and System Engineering, Rensselaer Polytechnic Institute, Troy, NY 12180, USA
}

\author{Boyang Zhang}
\affiliation{ 
Department of Electrical and Computer Engineering, University of Wisconsin at Madison, Madison, WI 53706, USA
}

\author{Tsung-Wei Huang}
\affiliation{ 
Department of Electrical and Computer Engineering, University of Wisconsin at Madison, Madison, WI 53706, USA
}

\author{Zhaoran Huang}
\affiliation{ 
Department of Electrical, Computer, and System Engineering, Rensselaer Polytechnic Institute, Troy, NY 12180, USA
}

\author{Duane S. Boning}
\affiliation{ 
Department of Electrical Engineering and Computer Science, Massachusetts Institute of Technology, Boston, MA 02139, USA
}

\author{Yu Yao}
\affiliation{ 
School of Electrical, Computer and Energy Engineering, Arizona State University, Tempe, AZ 85281, USA%
}

\author{Jiaqi Gu}
\affiliation{ 
School of Electrical, Computer and Energy Engineering, Arizona State University, Tempe, AZ 85281, USA%
}%
\email{jiaqigu@asu.edu.}

\date{\today}%

\input{doc/1_abs}

\maketitle

\input{doc/2_intro}

\input{doc/3_prelim}

\input{doc/4_algo}

\input{doc/5_exp}
\input{doc/6_conclu}

\section*{References}
\vspace{-20pt}
\bibliography{./ref/total}%
\newpage
\clearpage
\input{doc/7_appendix}

\end{document}

%% file: packages.tex
\usepackage[utf8]{inputenc} %
\usepackage[T1]{fontenc}    %
\usepackage[colorlinks,
    linkcolor=citelightblue,citecolor=citelightblue,urlcolor=citelightblue]{hyperref}       %
\usepackage{booktabs}       %
\usepackage{soul}
\usepackage[table,xcdraw]{xcolor}
\usepackage{graphicx}
\usepackage{threeparttable}
\usepackage{multirow}
\usepackage{amsmath}
\usepackage{bm}
\usepackage{bbm}
\usepackage{amsthm}
\usepackage{enumitem} %
\usepackage[subrefformat=parens,labelformat=parens]{subfig}

\usepackage{wrapfig}
\usepackage[ruled, vlined, linesnumbered, noend]{algorithm2e}
\usepackage{amssymb} %

\SetCommentSty{mycommfont}

\SetKwComment{TriangleComment}{$\blacktriangleright$~}{} %
\usepackage{pifont}
\usepackage{bm}

%% file: macro.tex
\definecolor{citecolor}{RGB}{34,139,34}
\definecolor{mydarkblue}{rgb}{0,0.08,1}
\definecolor{mydarkgreen}{rgb}{0.02,0.6,0.02}
\definecolor{mydarkred}{rgb}{0.8,0.02,0.02}
\definecolor{mydarkorange}{rgb}{0.40,0.2,0.02}
\definecolor{mypurple}{RGB}{111,0,255}
\definecolor{myred}{rgb}{1.0,0.0,0.0}
\definecolor{mygold}{rgb}{0.75,0.6,0.12}
\definecolor{myblue}{rgb}{0,0.2,0.8}
\definecolor{mydarkgray}{rgb}{0.,0.2,0.2}

\definecolor{lightred}{RGB}{255,235,235}
\definecolor{lightgreen}{RGB}{235,255,235}
\definecolor{lightblue}{RGB}{235,235,255}
\definecolor{citelightblue}{RGB}{49,164,222}
\definecolor{lightcyan}{RGB}{235,255,255}
\definecolor{lightmagenta}{RGB}{255,235,255}
\definecolor{lightyellow}{RGB}{255,255,235}

\definecolor{qxkcolor}{RGB}{215,235,255}
\definecolor{softmaxcolor}{RGB}{230,235,255}
\definecolor{probxvcolor}{RGB}{255,255,235}

\definecolor{topkcolor}{RGB}{255,235,235}
\definecolor{zecolor}{RGB}{255,255,235}
\definecolor{dynacolor}{RGB}{235,255,255}

\definecolor{reviewcolor}{RGB}{0,0,200}

\newcommand{\calO}{\mathcal{O}}

\newcommand{\calL}{\mathcal{L}}

\newcommand{\calT}{\mathcal{T}}

\newcommand{\eps}{\vec{\epsilon}}
\newcommand{\curl}{\nabla\times}

\renewcommand{\vec}[1]{\boldsymbol{#1}} 

\DeclareMathOperator*{\argmin}{argmin}

\theoremstyle{plain}

\theoremstyle{definition}

\newcommand{\vh}{\mathbf{h}}
\newcommand{\vb}{\mathbf{b}}

%% file: doc/1_abs.tex
\begin{abstract}
Diffractive Optical Neural Networks (DONNs) harness the physics of light propagation to perform speed-of-light, efficient analog computation, which has demonstrated diverse applications in artificial intelligence (AI) inference and signal processing.
Advances in nanophotonic fabrication and metasurface-based wavefront engineering have opened new pathways to realize high-capacity DONNs across various spectral regimes.
Training such DONN systems to determine the metasurface structures remains a challenging problem.
Heuristic methods are fast but oversimplify metasurfaces as element-wise phase masks or convolution units, often resulting in physically unrealizable designs and significant performance degradation.
Accurate simulation-in-the-loop training methods directly optimize a physically implementable metasurface using adjoint inverse design methods during end-to-end DONN training, but are inherently computationally prohibitive and unscalable.
To address the DONN training challenge in a physically feasible and scalable manner, we propose a spatially decoupled, progressive training scheme, \name.
For the first time, we formulate DONN training as a partial differential equation (PDE)-constrained learning problem, where metasurface responses are relaxed into freely-trainable transfer matrices with a banded structure. 
We then progressively enforce physical constraints through alternating transfer matrix training and adjoint inverse design. 
It eliminates the need for costly PDE solving per training iteration while ensuring physical realizability of the final metasurface design.
To further alleviate the inverse design runtime bottleneck, we introduce a physics-inspired, spatially decoupled inverse design strategy inspired by the natural locality in field interaction and field smoothness due to diffraction. 
We partition the metasurface into independently solvable patches and optimize the transfer matrix of each sub-region in parallel, followed by system calibration to restore global consistency.
We evaluate our \name scheme across a range of DONN training tasks.
Compared to state-of-the-art methods, \name achieves digital-comparable accuracy while being $\text{1825}\times$ faster than simulation-in-the-loop approaches due to the patched transfer matrices probing and much fewer iterations of inverse design.
Grounded in a physics-inspired optimization, \name bridges the gap between abstract DONN models and physical metasurface designs and paves the way for scalable training of meta-optical neural systems with guaranteed physical feasibility and high performance.

\end{abstract}

%% file: doc/2_intro.tex
\section{Introduction}
\vspace{-10pt}
\label{sec:Intro}
The rising demand for fast, energy-efficient inference in AI systems has spurred strong interest in unconventional computing paradigms.
Diffractive optical neural networks (DONNs) have emerged as a promising solution, leveraging the inherent parallelism, low-latency, and low-power characteristics of light propagation to perform analog computation at the speed of light~\cite{lin2018all, luo2022computational, mengu2022classification, mengu2022all, li2022polarization, Li_DiffAna_2023, bai2024pyramid, mengu2023snapshot,Li_lightridge_2024, wu2019neuromorphic, zhou2023physics, li2022rubikonns, chen2022physics, li2022physics, chen2022complex, xiang2022knowledge, tseng2021neural, colburn2019optical, qian2020performing, luo2019design, chen2021diffractive, luo2022metasurface, gu2024direct, yu2025all, tang2024optical,Hu2024Diffractive}.
One emerging technology to implement compact, flat optical field manipulation in DONNs is optical metasurfaces. 
Metasurfaces are engineered planar optical elements, typically composed of arrays of subwavelength-scale structures known as meta-atoms. 
These meta-atoms allow the metasurface to precisely modulate the phase and amplitude of incident light. This modulation, combined with free-space light diffraction in homogeneous media, enables the performance of mathematical operations analogous to those in artificial neural networks. %

Recent advances in nanofabrication and wavefront engineering have empowered DONNs with considerable design flexibility, enabling high-capacity implementations across a broad range of wavelengths~\cite{Ji2023Recent}.
These developments have led to successful demonstrations of DONNs in diverse tasks such as image processing, AI inference, scientific computing, sensing, and imaging.
However, realizing their full potential presents significant difficulties: training DONNs and designing physically feasible metasurface devices remain notable challenges, rooted in the complexities of optical physics. 

Traditional approaches to DONN design fall broadly into two categories: \emph{heuristic} approximation methods and \emph{simulation-in-the-loop} training~\cite{khoram2019nanophotonic}.
Heuristic methods simplify the metasurface as a set of element-wise phase masks or convolution kernels, often bypassing rigorous full-wave optical simulations by assuming local periodic behavior.
These methods translate the desired phase profile into a metasurface layout using a pre-simulated look-up table (LUT) that maps each target phase shift to a corresponding meta-atom design.
Though computationally efficient, this approach often neglects important inter-element optical interactions and relies heavily on the local periodic approximation (LPA)~\cite{wu2019neuromorphic}. 
The LPA assumes a smooth near-field response, a condition that might hold in some lens designs but is frequently violated in realistic DONNs where strong spatial variations are common.
As a result, heuristic methods often produce target phase profiles or transfer matrices that are not physically realizable.
When these idealized designs are projected onto feasible metasurface structures, the mismatch leads to significant deviations in optical behavior and degraded DONN performance.
This disconnect between abstract models and physical reality has been a persistent challenge. 
Efforts to mitigate this issue, e.g., by applying smoothness regularization~\cite{Li_lightridge_2024,zhou2023physics} to phase profiles to ease hardware design, have generally been inadequate. 
While smoother profiles are simpler to implement, they significantly limit DONN expressivity and do not ensure true physical feasibility, often leading to impractical designs. %

In contrast, simulation-in-the-loop training\cite{khoram2019nanophotonic} embeds metasurface optimization directly within the training loop by leveraging adjoint methods. These methods compute analytical gradients via forward and adjoint full-wave electromagnetic simulations.
By parameterizing the DONN trainable weights as physical metasurface design variables, such as meta-atom widths, this approach effectively enforces physical feasibility at every training iteration. %
However, the demanding requirement of solving Maxwell’s equations for each metasurface at each iteration makes this approach prohibitively expensive and fundamentally unscalable to the large, high-capacity optical systems needed for advanced AI tasks. %

To address the feasibility and scalability bottlenecks in DONN training, we propose \name, a physics-aware optimization scheme that progressively enforces Maxwell PDE constraints during training.
Unlike prior heuristic or simulation-in-the-loop approaches, \name alternates between relaxed DONN training and inverse metasurface projection, ensuring that learned responses remain physically realizable while maintaining efficient design space exploration.

To further reduce the significant computational cost, we develop a patch-wise metasurface simulation strategy. This strategy effectively exploits the locality of meta-atom interactions and inherent spatial frequency limits due to diffraction. 
This enables scalable metasurface inverse design with near-linear complexity, facilitating the design of physically implementable and scalable, large-capacity DONNs. %

Our main contributions can be summarized as follows:
\begin{itemize}[leftmargin=*]
\setlength{\itemindent}{0.5em}
\vspace{-5pt}
    \item We analyze the core limitations of existing DONN training methods, which suffer from modeling inaccuracies and the prohibitive simulation cost. 
    We introduce \name, a scalable PDE-constrained DONN inverse optimization framework that ensures physical feasibility while drastically reducing simulation cost. %
    \vspace{-5pt}
    \item \textbf{Spatially-Decoupled Metasurface Simulation}: \name exploits the locality of meta-atom interactions to divide the metasurface into patches, enabling scalable, parallel simulation and reducing complexity from cubic to near-linear.%
    \vspace{-5pt}
    \item\textbf{Diffraction-Inspired Transfer Matrix Sketching}: By leveraging the natural smoothness of diffracted fields, we introduce a pulse-based sketching technique that accelerates transfer matrix probing without sacrificing accuracy.
    \vspace{-5pt}
    \item \textbf{Progressive PDE-Constrained Learning}: 
    \name features a novel alternating scheme that interleaves relaxed DONN training with adjoint-based projection onto the Maxwell-constrained subspace, ensuring physical realizability throughout training while maintaining design flexibility and expressivity. 
    \vspace{-5pt}
    \item On multiple DONN benchmarks, \name achieves up to \textbf{63.88\%} higher accuracy and delivers an \textbf{1825$\times$} speed-up compared to state-of-the-art simulation-in-the-loop training methods because of the patched transfer matrices probing and much fewer iterations of inverse design, bridging the gap between analytical DONN training and physically realizable meta-optic hardware. %
\end{itemize}

%% file: doc/3_prelim.tex
\vspace{-10pt}
\section{Background}
\label{sec:prelim}
\vspace{-10pt}
This section introduces the fundamental concepts and existing methods. 
First, we describe the physical principle of DONNs. 
Then, we explain the architecture of a hybrid DONN. Finally, we review state-of-the-art (SoTA) DONN training methods and pinpoint their limitations.

\vspace{-10pt}
\subsection{Transfer Matrix Description of DONN Responses}
\label{sec:light_propagation}
\vspace{-10pt}
Light propagation in DONNs involves two linear physical processes: free-space diffraction and metasurface-based light modulation, as illustrated in Fig.~\ref{fig:hybrid_DONN}.

\noindent\textbf{Free-Space Diffraction.}~
Light diffraction in a homogeneous medium can be precisely modeled using a Green’s function formulation, which provides an analytical near-to-far field transformation (N2FF). 
This representation naturally involves Hankel functions to describe cylindrical wave propagation.

\noindent\textbf{Metasurface-based Modulation.}~
Metasurfaces modulate the phase and amplitude of incident light through engineered subwavelength structures. 
The response of a metasurface to incoming light is governed by Maxwell’s equations, which in the frequency domain reduce to solving a linear system of the form $A x = b$, where $A$ is the discretized Maxwell operator, $x$ is the electromagnetic field, and $b$ is the excitation source.

We define two types of transfer matrices: the diffraction matrix $U(z)$, which captures the field propagation over a distance $z$ in a homogeneous medium, and the modulation matrix $\mathcal{T}(\epsilon)$, which encodes the response of a metasurface parameterized by its permittivity distribution $\epsilon$.
As shown in Fig.~\ref{fig:hybrid_DONN}, the light propagation process in a $k$-layer DONN, comprising alternating layers of modulation and diffraction, can be described using a sequence of linear operators.
\begin{equation}
    \small
    \label{eq:DONN}
    \begin{aligned}
            Y=|H_{out}|^2, ~~H_{out}=U_{k+1}\calT_k(\epsilon_k)\cdots \calT_3(\epsilon_3)U_2\calT_1(\epsilon_1)U_1 x
    \end{aligned}
\end{equation}
where the input data will be first encoded to the intensity or phase of complex-valued input light field $x$.
In prior work, based on the LPA assumption, a metasurface with $n$ meta-atoms is often modeled as an element-wise modulation plate with diagonal response $\calT=\text{diag}([A_1e^{j\phi_1}, \cdots, A_n e^{j\phi_n}])$.
At the end, the light will be converted to electrical signals via photodetector arrays (or image sensors) with a square function applied $Y=|\cdot|^2$.
The metasurface transfer matrix $\mathcal{T}(\epsilon)$ can be numerically obtained by stimulating the system with a set of orthonormal basis vectors, typically the identity matrix, to capture its full impulse response, akin to classical system identification methods in linear algebra and signal processing.

\input{figtex/fig_hybrid_DONN}
\vspace{-10pt}
\subsection{Hybrid Diffractive Optical Neural Network}
\vspace{-10pt}
It is a promising trend to hybridize DONN with digital neural networks for adaptive AI inference\cite{yin2024reconfigurable, wei2024spatially, liu2025ultra}.
Figure~\ref{fig:hybrid_DONN} illustrates the architecture of a hybrid DONN with an optical feature extraction module and a lightweight digital neural head. 
An input signal is encoded into the phase of light and injected into the optical system via waveguides. 
As the light propagates through multiple layers, it undergoes metasurface-based modulation, $\mathcal{T}_i$, and free-space diffraction, $U$, forming a characteristic intensity distribution on the detector plane. This intensity pattern is spatially partitioned into disjoint regions, and the light intensities within each region are integrated to produce output values for different channels. Consequently, the optical module functions similarly to a convolutional layer with a single input channel and $n$ output channels. The intensity integration process introduces a built-in nonlinearity.

\vspace{-10pt}
\subsection{Overview and Analysis of DONN Training Methods}
\label{sec:related_work}
\vspace{-10pt}
In general, the DONN training can be separated into two classes: the offline training and \emph{in-situ} training. 
\vspace{-10pt}
\subsubsection{Offline Training}
\vspace{-10pt}
\noindent\textbf{Heuristic DONN Training Methods.}
Heuristic approaches aim to accelerate DONN training by simplifying the underlying physics, avoiding costly transfer matrix simulations during optimization~\cite{lin2018all, luo2022computational, mengu2022classification, mengu2022all, li2022polarization, Li_DiffAna_2023, bai2024pyramid, mengu2023snapshot,Li_lightridge_2024, wu2019neuromorphic, zhou2023physics, li2022rubikonns, chen2022physics, li2022physics, chen2022complex, xiang2022knowledge, tseng2021neural, colburn2019optical, qian2020performing, luo2019design, chen2021diffractive, luo2022metasurface, gu2024direct, yu2025all, tang2024optical}. 
Instead, they train with overly simplified metasurface models and implement the physical design post hoc. 
While computationally efficient, these methods often suffer from a mismatch between the trained model and its physical realization.

A widely used heuristic method models the metasurface modulation as element-wise phase shifts, treating its transfer matrix $\mathcal{T}$ as a diagonal matrix with unit amplitude. 
It assumes that each meta-atom operates independently with periodic boundary conditions. 
After training, the desired phase profile is mapped to a physical metasurface design by assembling meta-atoms from a pre-simulated library, i.e., look-up table (LUT).
However, this diagonal assumption neglects inter-element optical coupling, making the resulting $\mathcal{T}$ physically unrealizable. 
In practice, the optical response of a meta-atom depends not only on its own geometry but also on the configuration of its neighbors. 
Deviations from the idealized periodic setting, such as variation in adjacent meta-atom sizes or spacings, violate the LPA and introduce substantial performance degradation during deployment.

Another heuristic method attempts to account for meta-atom interactions by modeling the metasurface modulation as a convolution operation~\cite{wu2019neuromorphic}, still relying on the LPA. 
In this approach, the transfer matrix is parameterized as a learned convolution kernel, from which the physical metasurface design is directly derived.
While this captures some local coupling effects, it remains an oversimplification of the underlying optical physics. 
The reliance on LPA still limits accuracy, especially in regimes where strong inter-element interference or non-periodic layouts cause the approximation to break down.

\noindent\textbf{Simulation-based Training Methods.}
To ensure physical fidelity, simulation-in-the-loop methods embed full-wave electromagnetic simulations directly into the DONN training process~\cite{khoram2019nanophotonic}. 
At each iteration, the metasurface modulation matrix $\mathcal{T}$ is simulated by solving Maxwell equations, and gradients are computed via the adjoint method, which requires both forward and adjoint simulations.
While this approach provides accurate modeling and guarantees physical feasibility throughout training, its computational cost scales poorly with system size. 
The need to solve large-scale PDEs at every step makes it prohibitively expensive for large or multi-layer metasurface designs, severely limiting its scalability for practical DONN training.
Another simulation-based method~\cite{sun2024multimode} constructs a response LUT by exhaustively simulating all possible device configurations within a discretized design space. The classification accuracy of each design is then evaluated, and the one achieving the highest performance is selected. However, this approach is inherently limited by the finite and discrete nature of the design space, which enables brute-force enumeration but restricts scalability to continuous or high-dimensional design problems. Moreover, the method lacks generalizability to other types of meta-optical structures beyond the predefined library.
\input{tables/tab_baselineComplexity}

Table~\ref{tab:baselineComplexity} compares the computational complexity of different offline DONN training paradigms, decomposed into two components: training cost and design cost. The training cost is proportional to the number of epochs and dataset size across all methods. Heuristic methods achieve low design cost by relying on precomputed LUTs, resulting in constant-time design after training. In contrast, simulation-in-the-loop approaches tightly couple training and physical simulation, incurring high per-iteration cost due to repeated full-wave PDE solves, leading to overall complexity that scales poorly with system size. 
\name~offers a favorable trade-off by decoupling training from simulation. Its design cost scales linearly with metasurface size, thanks to the use of localized, patched transfer matrix probing. As a result, \name~retains physical fidelity while remaining computationally tractable. In Appendix~\ref{sec:vis_larger_sys}, we compare the runtime needed between the proposed \name and vanilla full-size simulation for different sizes of metasurfaces.

\vspace{-10pt}
\subsubsection{In-situ Training}
\vspace{-10pt}
Besides offline training, another category is on-chip/\emph{in-situ} training that tunes modulation signals, such as voltage or heat, applied to fabricated meta-optical structures by directly measuring physical outputs as feedback to update the system parameters, thereby improving post-fabrication performance~\cite{cheng2024multimodal, zhou2020situ, xue2024fully, NP_Science2023_Pai}. \emph{In-situ} training is typically used in post-fabrication calibration and online fine-tuning, which is complementary to offline training. 
This work focuses on developing an offline training framework that produces high-performance, fabrication-ready designs.
\vspace{-13pt}
\subsection{Metasurface Design Methods}
\vspace{-10pt}
Various methods have been proposed for metasurface design. 
A widely adopted approach is the LUT-based method~\cite{aieta2015multiwavelength, chen2018broadband}, where design parameters are systematically swept and their optical responses are recorded via numerical simulations to construct a meta-atom library. 
Given a desired phase or amplitude profile, suitable meta-atoms are then selected by querying the LUT for entries that best match the target response. 
However, this method typically neglects the electromagnetic interactions between neighboring meta-atoms, which can result in \textit{significant performance degradation} after fabrication and deployment. 
The second class of metasurface design methodologies falls under zero-order/gradient-free optimization methods, such as genetic algorithms~\cite{chen2023polychromatic, fan2020phase, jafar2018adaptive}. 
These approaches do not require gradient information and can perform a global search to avoid local optima. 
However, zero-order methods cannot scale to high-dimensional optimization and typically demand a large number of full-wave simulations to sufficiently explore the vast design space, resulting in high computational cost. 
This \textit{inefficiency} makes them less suitable for scenarios that require repeated or large-scale optimization. 
Adjoint method-based inverse design~\cite{mansouree2021large, zhang2022inverse, oh2022adjoint, phan2019high} represents another class of methodologies, in which the gradient of the objective function w.r.t. the design variables can be computed analytically through only two full-wave simulations, one forward and one adjoint. This enables \textit{highly efficient optimization} in high-dimensional design spaces. 
However, the adjoint-based method relies on multiple computationally intensive full-wave simulations, which are prohibitively time-consuming on large devices. 
Previous work~\cite{phan2019high} proposed a disjoint patch-based strategy to reduce the computational complexity of metasurface optimization from polynomial to linear. The first key distinction from our approach lies in the treatment of boundary conditions. While both methods simulate small patches independently, our use of weak probing stimuli centered within each patch ensures that the field near patch boundaries is negligible, thereby mitigating boundary artifacts. In contrast,~\cite{phan2019high} applies excitation across the entire patch, making it more sensitive to boundary-induced errors.
The second key difference lies in the answer to the following question:\\
\noindent{\underline{\textit{Is DONN training simply an inverse design problem?}}} 
DONN training is fundamentally more challenging than metasurface inverse design tasks. 
Inverse design solves a deterministic optimization problem for a specific figure-of-merit. 
In contrast, DONNs must learn metasurface designs on thousands of input-label pairs on the training dataset and generalize to the test set, making the problem a stochastic, data-driven learning task.
Every forward pass per training iteration imposes strict Maxwell PDE constraints. 
This makes the naive use of full-wave simulations in every iteration, as required by simulation-in-the-loop methods, computationally infeasible at scale.
We claim that DONN training can be viewed as a strict superset of inverse design, requiring \textbf{fundamentally different methodologies to balance trainability, convergence, and efficiency.}

%% file: figtex/fig_hybrid_DONN.tex
\begin{figure}
    \centering
    \includegraphics[width=\columnwidth]{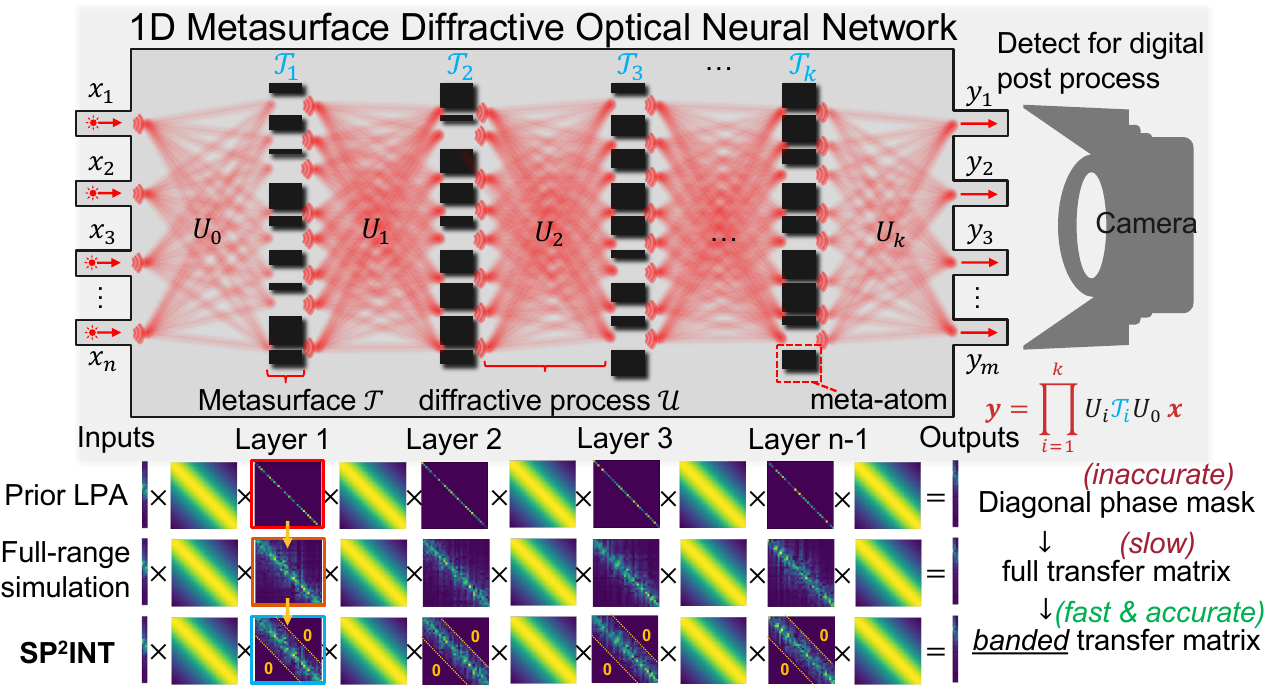}
    \caption{DONN with multi-layer metasurfaces can be modeled as cascaded transformation $\calT$ and diffraction $U$.
    Metasurface transfer matrix modeling comparison among 3 methods.
    Our \name uses banded transfer matrix probing for fast and accurate modeling.
    }
    \label{fig:hybrid_DONN}
\end{figure}

%% file: tables/tab_baselineComplexity.tex
\begin{table}[]
\centering
\caption{DONN training complexity comparison among different methods. 
$E$ denotes the number of training epochs, $N$ is the size of the training dataset, $B$ is the inverse design iteration budget for each training epoch ($B\ll N$), and $n$ represents the spatial dimension of the metasurface.
}
\resizebox{0.85\columnwidth}{!}{%
\begin{tabular}{|cc|c|}
\hline
\multicolumn{2}{|c|}{Offline DONN Training Methods}                                           & \multicolumn{1}{l|}{Algorithmic Complexity} \\ \hline
\multicolumn{1}{|c|}{\multirow{2}{*}{Heuristics-based}} & LPA\cite{luo2022metasurface}                     & $\mathcal{O}(EN+1)$                            \\ \cline{2-3} 
\multicolumn{1}{|c|}{}                           & Convolution\cite{wu2019neuromorphic}                                 & $\mathcal{O}(EN+1)$                            \\ \hline
\multicolumn{1}{|c|}{Simulation-based}                 & Simulation-in-the-loop\cite{khoram2019nanophotonic} & $\mathcal{O}(ENn^3)$       \\ \hline\hline
\multicolumn{1}{|c|}{Simulation-based}                 & \name & $\mathcal{O}(EN+EBn)$       \\ \hline
\end{tabular}
}
\label{tab:baselineComplexity}
\end{table}

%% file: doc/4_algo.tex
\vspace{-10pt}
\section{Proposed DONN Training Scheme \name}
\vspace{-10pt}
We aim to train hybrid DONNs under physically realistic conditions, where light propagation and metasurface behavior obey Maxwell's equations. To this end, we formulate DONN training as a PDE-constrained subspace optimization problem, which presents two major challenges: ensuring physical realizability and achieving scalability. To address both, we introduce \name, a progressive training framework that alternates between relaxed DONN learning and adjoint-based inverse design. \name~integrates two key innovations: \textbf{progressive soft projection}, which gradually enforces physical constraints while enabling efficient design space exploration, and \textbf{patched transfer matrix probing}, which enables scalable modeling of large metasurfaces without resorting to full-system simulations.

\vspace{-10pt}
\subsection{Problem Formulation}
\vspace{-10pt}
Instead of being formulated as a deterministic optimization problem as in conventional inverse design, hybrid DONN training can be formulated as a stochastic learning problem:
\begin{equation}
\small
\label{eq:Formulation}
\begin{gathered}
\eps^*, w^* = \argmin _{\eps, w} \mathbb{E}_{(\mathbf{x}, y) \sim \mathcal{D}}\left[\mathcal{L}\left(g_{w}\circ f(\mathbf{x}; \eps), y\right)\right],\\
f(\mathbf{x};\eps)=|\vh|^2,\\
\text{s.t.}~~\big(\curl(\eps_{\text{tot}}^{-1}\curl)-\omega^2\mu_0\epsilon_0\big)\vh=\vb(\mathbf{x}) ~~\rightarrow~~ A(\eps)\vh=\vb(\mathbf{x}).
\end{gathered}
\end{equation}
Here, $f(\mathbf{x}; \eps)$ models the diffractive optical system, its output being the light intensity detected on photodetector arrays (for TM polarized light, this intensity is proportional to $|\vh|^2$, where $\vh$ is the magnetic field). 
The function $g_w(\cdot)$ represents the subsequent digital processing head, encompassing operations like normalization, biasing, and the final digital classification layers, with $w$ as its learnable variables. 
The primary trainable parameters for the optical system are the $k$ metasurface structures, characterized by their permittivity distributions $\eps=(\epsilon_1,\cdots,\epsilon_k)$. 
Input data and label pairs $(\mathbf{x}, y)$ are drawn from the training dataset $\mathcal{D}$, and $\mathcal{L}$ is the chosen loss function.
The magnetic field $\vh$ follows Maxwell's equation, which sets a challenging Maxwell PDE constraint to the optimization problem.
$\eps_{\text{tot}}$ describes the entire optical system, including multiple cascaded metasurfaces and their spacing cladding.
For simplicity, we use a linear equation for the PDE constraint $A(\eps)\vh=\vb(\mathbf{x})$, where $\vh$ is the vectorized magnetic field and $\vb(\mathbf{x})$ is the vectorized input source related to $\mathbf{x}$.

This PDE-constrained formulation dictates that each optimization iteration requires solving the forward Maxwell equation to obtain $\vh$, followed by solving an adjoint Maxwell equation to efficiently compute the gradients $\frac{d\calL}{d\eps}$ with respect to the metasurface parameters:
\begin{equation}
    \small
    \begin{aligned}
    A(\eps)^{\top}\vh_{\text{adj}}=-\frac{\text{d}\calL}{\text{d} \vh},~~~\frac{\text{d}\calL}{\text{d}\eps}=-\Re(\vh_{\text{adj}}^\top \vh).
    \end{aligned}
\end{equation}

However, rigorous full-wave simulation of the entire optical system by solving the PDE is prohibitively time-consuming, making it impractical to embed this simulation within the outer training loop.

\vspace{-13pt}
\subsection{Understanding the Difficulty in Designing Implementable DONNs} %
\vspace{-10pt}
The formulation in Sec.~\ref{sec:prelim} formulates DONN training as a PDE-constrained optimization problem. In this section, we first examine the fundamental nature of this constraint: how Maxwell’s equations govern the physical realizability of the optical system, and then identify the key optimization challenges and computational cost it introduces. These insights motivate the design of \name.

\vspace{-10pt}
\subsubsection{The Essence of the PDE Constraint}
\vspace{-10pt}
The core constraint in Eq.~\eqref{eq:Formulation} requires that the electromagnetic field $\vh$ throughout the diffractive system strictly satisfies Maxwell’s equations, given a metasurface design $\eps$. For a DONN composed of $k$ cascaded metasurfaces, we analyze this constraint by modeling the system as a sequence of modulation and diffraction operations, as supported by the Transfer Matrix Method (TMM), which naturally enables a layer-wise decomposition of wave propagation, shown in Fig.~\ref{fig:hybrid_DONN}.
Specifically, each metasurface with permittivity $\epsilon_i$ modulates an incident field $\mathbf{x}_i$ according to a local PDE constraint:
\begin{equation}
\label{eq:layer_maxwell}
\small
\left(\nabla \times\left(\epsilon_i^{-1} \nabla \times\right)-\omega^2 \mu_0 \epsilon_0\right) \mathbf{h}_i=\mathbf{b}(\mathbf{x}_i)\Rightarrow A(\epsilon_i)\mathbf{h}_i=\mathbf{b}(\mathbf{x}_i), \text{for } i\in \mathbb{N^+}.
\end{equation}
Each local interaction defines a modulation transfer matrix $\mathcal{T}_i(\epsilon_i)$ that maps the input to the output field near the metasurface. Combined with free-space diffraction matrices $U_i$, the full end-to-end system response becomes:
\begin{equation}
\label{eq:overall_transfer_matrix_prod}
f(\mathbf{x}_{\text{in}}; \eps) = \left|\left( \prod_{j=1}^{k} U_j \mathcal{T}_j(\epsilon_j) \right) U_0 \mathbf{x}_{\text{in}} \right|^2 = |\mathbf{h}_{\text{out}}|^2.
\end{equation}
To ensure that the trained DONN system is physically realizable, each $\mathcal{T}_i(\epsilon_i)$ must correspond to the response of an actual metasurface governed by Eq.~\eqref{eq:layer_maxwell}.

This leads to a fundamental question: \textit{Is an arbitrary transfer matrix $\mathcal{T}$ physically implementable by some passive metasurface $\epsilon$?} The answer is unequivocally no. For instance, a non-unitary matrix for a lossless system or a matrix implying gain would violate energy conservation and thus cannot represent any physically passive optical component. \textbf{Consequently, training a DONN that adheres to Maxwell's equations inherently becomes a constrained optimization problem, where the learned transfer matrices $\widehat{\mathcal{T}}_i(\epsilon_i)$ must reside within the subspace of physically implementable transformations}, which we refer to as the \textit{implementable subspace}.

\vspace{-10pt}
\subsubsection{Training Difficulties} 
\label{sec:training_difficulties}
\vspace{-10pt}
As discussed in Sec.~\ref{sec:related_work}, the primary challenge lies in the oversimplification of transfer matrices by heuristic methods, which lead to unimplementable responses and subsequent performance degradation, while rigorous simulation-in-the-loop training remains prohibitively expensive. 
We identify \textbf{three key difficulties} that hinder the design of physically implementable and high-performance DONNs:
\ding{202} \textbf{Characterizing the Valid Subspace:} 
The intricate optical physics, encompassing wave interference and complex light-matter interactions, makes a precise analytical characterization of the implementable subspace exceptionally challenging.
Without clear boundaries for this valid subspace, it is difficult to guide the optimization effectively.
\ding{203} \textbf{Prohibitive Simulation Cost:} The coupling between each transfer matrix $\mathcal{T}_i$ and its underlying physical permittivity distribution $\epsilon_i$ is dictated by Maxwell's PDEs. 
Directly enforcing physical realizability by repeatedly solving PDEs for each metasurface design during training is computationally formidable. 
The cost of such full-wave simulations increases drastically with large metasurface sizes (e.g., $\calO(n^3)$ in FDFD), rendering this approach impractical for large systems and severely hindering the scalability of DONN training for realistic applications.
\ding{204} \textbf{Non-Convex Optimization Landscape:} 
The subspace of physically implementable transfer matrices $\mathcal{T}_i$ is highly non-convex. 
Optimizing within such a complex, non-convex landscape significantly increases the risk of the training process prematurely converging to bad local optima. 
This, in turn, curtails effective exploration of the vast design space and impedes the discovery of high-expressivity metasurface designs.

To effectively address these challenges, several fundamental \textbf{questions} must first be answered:

\noindent\textit{Q1. How can we impose the PDE constraints to restrict transfer matrices within the implementable subspace?}
As discussed earlier, PDE constraints are equivalent to enforcing the physical validity of trained metasurface transfer matrices. 
Generally, three strategies are common for imposing constraints in optimization: penalty, reparameterization, and projection. 
The penalty method relies heavily on incorporating complex physical priors, like enforcing energy conservation, yet it often fails to guarantee strict constraint satisfaction. 
Reparameterization can be pursued through numerical simulation or heuristic modeling. The former is prohibitively time-consuming for optimization. The latter lacks a clear and accurate formulation to map metasurface design variables to their transfer matrices. Previous heuristic methods, which often simplify the transfer matrix to representations such as a pure diagonal phase-shift matrix or a convolution kernel, have shown significant performance degradation, further underscoring the difficulty and inaccuracy of this approach.
\textbf{Therefore, projection emerges as a more promising and practical approach to enforce PDE constraints.}

\noindent\textit{Q2. How can we avoid per-iteration, unscalable PDE solving?}
The core PDE constraint, as previously established, is implicitly imposed by Maxwell’s equations over the entire DONN system. 
Leveraging TMM allows for the decomposition of this global constraint into $k$ \emph{layer-wise} problems. 
While TMM allows us to decompose global PDE constraints into layer-wise subproblems, each layer still requires expensive forward/adjoint simulations on the whole metasurface to obtain gradients w.r.t. its permittivity.
To make training scalable, we must \textbf{decouple field evaluation $\vh_i$ from design updates on $\epsilon_i$ and solve Maxwell PDEs only when necessary.}

\noindent\textit{Q3. How to perform projection to balance exploration and efficiency?}
Since metasurface inverse design is inherently a slow and difficult binary optimization problem, effective training requires a carefully scheduled projection strategy.
Hard projection strictly enforces feasibility but limits exploration and increases simulation cost, while soft projection improves efficiency and optimization flexibility but risks constraint violation, thus fails to converge.
\textbf{To ensure fast convergence and maintain physical realizability, we need to balance these trade-offs through progressive projection that gradually tightens constraints during training.}

To handle the aforementioned difficulties, we propose \name, a spatially decoupled physics-inspired progressive inverse optimization for scalable, PDE-constrained DONN training, demonstrated in Alg.~\ref{alg:prog_proj_train}, Alg.~\ref{alg:projection}, and Alg.~\ref{alg:patch_probe}

\vspace{-10pt}
\subsection{Proposed Decoupled Patched DONN Design Flow}
\label{sec:decoupled_training_flow}
\vspace{-10pt}
To avoid full-wave simulation for the entire optical system per learning iteration, we first apply the TMM to decompose the transfer function $f$ into multiple cascaded linear systems, i.e., $f(x;\eps)=|\prod_{i=1}^kU_i\calT_i(\epsilon_i)U_0x|^2$, where $U$ matrix is the diffraction process and $\calT_i(\epsilon_i)$ is the transfer matrix of the $i$-th metasurface.
$U$ is fixed once the diffraction distance is determined, which can be easily and precisely computed via near-to-far field projection.
In this way, we only need to obtain $\calT$ by simulating the metasurface's near-field response and analytically obtaining the system response using TMM.
To obtain the transfer matrix $\calT$, one can shine an identity matrix as a series of point sources through the metasurface and obtain the complex field on the near-field monitor, \begin{equation}
    \small
    \calT_i\in\mathbb{C}^{n\times n}=[\vh_1,\cdots\vh_n]=A(\epsilon_i)^{-1}B=A(\epsilon_i)^{-1}[\vb_1,\cdots,\vb_n],
\end{equation}
where the number of simulations $n$ depends on the discretization resolution.
\input{algo/algo_progressiveProj}

To efficiently tackle this problem, we propose a variable separation method to decouple the transfer matrix $\widehat{\calT}$ used in DONN training from the physical metasurface response $\calT$ that requires $n$ \emph{full-metasurface} simulations to extract.
\begin{equation}
    \small
    \label{eq:Decouple}
    \begin{aligned}
    \widehat{\calT}^*, w^* = \argmin_{\widehat{\calT}, w} \mathbb{E}&_{(\mathbf{x}, y) \sim \mathcal{D}}\left[\mathcal{L}\left(g_{w}\circ f(\mathbf{x}; \widehat{\calT}), y\right)\right],\\
\text{s.t.}~~~&A(\epsilon_i)\calT_i=B,~\forall i\in[k]\\
& \widehat{\calT}_i = \calT_i,~\forall i\in[k].
    \end{aligned}
\end{equation}
To solve this constrained optimization problem, the projected gradient descent can be adopted, as shown in Alg.~\ref{alg:prog_proj_train}, where we alternatively solve two subproblems.
The first subproblem is \ding{202}~\textbf{Relaxed DONN Training}:
\begin{equation}
    \small
    \label{eq:DONNTrain}
    \begin{aligned}
    \widehat{\calT}^{t+1}, w^{t+1} = \argmin_{\widehat{\calT}, w} \mathbb{E}&_{(\mathbf{x}, y) \sim \mathcal{D}}\left[\mathcal{L}\left(g_{w^t}\circ f(\mathbf{x}; \widehat{\calT}), y\right)\right].
    \end{aligned}
\end{equation}
Then we solve the second subproblem: \ding{203}~\textbf{Metasurface Inverse Design} by optimizing the metasurface permittivity $\eps$:

\begin{equation}
    \small
    \label{eq:Mapping}
    \begin{aligned}
    \eps^{t+1} = &\argmin_{\eps} \sum_{i=1}^k\|\calT_i-\widehat{\calT}_i^{t+1}\|_F^2,\\
    \text{s.t.}~~~&A(\epsilon_i)\calT_i=B,~\forall i\in[k],
    \end{aligned}
\end{equation}
as shown in Alg.~\ref{alg:projection}. 
The most time-consuming one to solve is the second inverse design problem, which involves multiple full-metasurface forward/adjoint simulations to find corresponding metasurface designs to match the given target matrix. To make matters worse, the computational complexity of these simulations scales rapidly with device size.
\input{algo/algo_proj}
\vspace{-10pt}
\subsection{Spatially-Decoupled Metasurface Simulation for Scalable Transfer Matrix Probing}
\label{sec:patch_probe_tm}
\vspace{-10pt}
Recall that in primal problem \ding{203}, we need to extract $\calT_i$ for all $k$ metasurfaces by solving $nk$ full-metasurface Maxwell equations in total, which is computationally expensive.
Based on a strong physics prior that meta-atom responses often have restricted non-local interaction, we can observe a clear \textbf{banded diagonal} structure on the transfer matrix $\calT$.
In other words, the light source shined on one meta-atom will not scatter to far-away locations in the near field with a large diffraction angle, such that we only need to simulate a small region and ignore all light transmission in the exterior region up to a certain acceptable approximation error as shown in Fig.~\ref{fig:patch_probe}.
\input{algo/algo_tmProbe}
Motivated by this observation, we propose a spatially-decoupled metasurface simulation approach by cutting a large metasurface into overlapping patches of $P$ meta-atoms and a stride of 1 atom.
The detailed algorithm is in Alg.~\ref{alg:patch_probe}.
In this way, we only need to simulate $n$ small size-$P$ patches, which significantly reduces the simulation cost when extracting the transfer matrix. 
\input{figtex/fig_patchProbe}
More importantly, the simulation cost now \textbf{scales linearly with the size of the metasurface}, transforming an otherwise cubic complexity into a much more scalable approach.

\vspace{-10pt}
\subsection{Progressive PDE-Constrained Training}
\label{sec:progressive_proj}
\vspace{-10pt}
\input{figtex/fig_progressiveProj}
To maintain the target matrix $\widehat{\mathcal{T}}_i$ within or close to the implementable subspace for better convergence, \name periodically projects each unconstrained matrix $\widehat{\mathcal{T}}_i$ to its nearest physically realizable counterpart $\mathcal{T}_i$ via adjoint inverse design.
A naive projection schedule that applies this projection only once per training epoch fails in practice. 
As shown in Fig.~\ref{fig:progressive_proj}, a full epoch of unconstrained training often causes $\widehat{\mathcal{T}}_i$ to drift too far from the implementable subspace, such that the subsequent projection step can no longer find a physically realizable metasurface that accurately recovers the target response. 
This results in severe performance drop and even divergence.
Increasing the projection frequency helps prevent $\widehat{\mathcal{T}}_i$ from drifting too far from the implementable subspace, making it more likely that a physically realizable design can be found. 
However, due to the highly non-convex nature of the implementable subspace, as discussed in Sec.~\ref{sec:training_difficulties}, overly frequent hard projection often causes the optimization to get stuck near the initialization without meaningful progress, shown in Fig.~\ref{fig:progressive_proj}.
To resolve this, we introduce a \textit{progressive PDE-constrained training} strategy. 
Rather than enforcing hard binarization from the start, we allow the inverse design module to initially project to relaxed, continuous-valued device patterns that lie near the implementable subspace. 
As training proceeds, we gradually tighten the binarization constraint, ensuring the final metasurface designs are both expressive and physically realizable.

\vspace{-10pt}
\subsection{Extra System-Level Fine-Tuning for Enhanced Optimality}
\label{sec:fine_tune_entire}
\vspace{-10pt}
Inspired by the concept of layer-wise NN knowledge distillation, where intermediate layer features are aligned besides matching the final output, we introduce an extra system-level fine-tuning stage. 
Besides matching the TM $\mathcal{T}_i$ of each individual metasurface to its corresponding target matrix $\widehat{\mathcal{T}_i}$ as in Eq.~\eqref{eq:Mapping}, we further calibrate the response of the entire optical system, in which light propagates through multiple stacked metasurfaces. 
This is achieved by optimizing the metasurface $\epsilon$ to minimize the system-level objective in Eq.~\eqref{eq:ResponseMatching}, employing the projection methods in Alg.~\ref{alg:projection}.
\begin{equation}
    \small
    \label{eq:ResponseMatching}
    \begin{aligned}
    \epsilon^{t+1}
    = \arg\min_{\epsilon} &\left\| \mathcal{T}_{\text{tot}} S - \widehat{\mathcal{T}}_{\text{tot}}^{t+1} S \right\|_F^2, \quad
    \mathcal{T}_{\text{tot}} = \prod_{i=1}^k U_i \mathcal{T}_i(\epsilon_i), \\
    \text{s.t.} &\quad
    A(\epsilon_i) \mathcal{T}_i = B, \quad \forall i \in [k].
    \end{aligned}
\end{equation}
In Eq.~\eqref{eq:ResponseMatching}, $\mathcal{T}_{\text{tot}}$ represents the actual total transfer matrix of the multi-layer physical system. The target total transfer matrix, $\widehat{\mathcal{T}}_{\text{tot}}^{t+1}$, is constructed from the layer-wise target modulation matrices $\widehat{\mathcal{T}}_i^{t+1}$ obtained from the DONN training stage shown in Eq.~\eqref{eq:DONNTrain}, typically as $\widehat{\mathcal{T}}_{\text{tot}}^{t+1} = \prod_{i=1}^k U_i \widehat{\mathcal{T}}_i^{t+1}$. The matrix $S$ consists of a set of one-hot probe stimuli.

This system-level objective allows us to exploit the inter-layer interactions between metasurfaces, enabling improved alignment with the desired global optical response and ultimately enhancing end-to-end performance.

%% file: algo/algo_progressiveProj.tex
\begin{algorithm}
\caption{Progressive Projected Training for Multi-Layer DONN}
\label{alg:prog_proj_train}

\KwIn{
Model $\mathcal{M}_{\widehat{\mathcal{T}}, w}$, use $\mathcal{M}$ for short; Training set $\mathcal{D}(x, y)$; Uniform metasurface init $\epsilon^0$; 
Initial and final binarization sharpness: $s_0$, $s_T$; 
Inverse design iteration budget per epoch $B$, number of inverse design iteration per projection $I$; 
Number of metasurface layers $K$; number of meta atoms in each metasurface $N$; patch size $P$
}
\KwOut{Trained model with implementable metasurfaces}

$\widehat{T^{0}} \leftarrow$ \texttt{ProbeTM}($\epsilon^0$, $s_T$, $P$, $N$)\tcp*{Sec.~\ref{sec:patch_probe_tm}}
\For{$\text{k} \leftarrow 1$ \KwTo $\text{K}$}{
$\mathcal{M}$\texttt{.SetTM}($\widehat{T^{0}}$, k) \tcp*{Init $k^{th}$ layer}
}
$\eps^0 \leftarrow$ [$\epsilon^0$] $\times K$ \tcp*{Init all K layers using $\epsilon^0$}
$n_p \leftarrow B / I$ \tcp*{Number of proj per epoch}
$b_p \leftarrow \texttt{len}(\mathcal{D}) / n_p$ \tcp*{Batch interval between proj}
$\mathcal{S} \leftarrow \texttt{Sched}(s_0, s_T, B)$ \tcp*{Define sharpness scheduler}

\For(\tcp*[f]{Sec.~\ref{sec:decoupled_training_flow}}){$\text{epoch } t \leftarrow 1$ \KwTo $n_{\text{epochs}}$}{
  $\eps^t\leftarrow\eps^{t-1}$\;
  \ForEach{batch index $i$, $(x, y)$ in $\mathcal{D}$}{
    \texttt{TrainStep}($\mathcal{M}$, x, y) \tcp*{Forward + loss + backward + optimizer}

    \If{$i \bmod b_p = 0$ \textbf{and} $i \neq 0$}{
      $\eps^t \leftarrow$ \texttt{Project}($\mathcal{M}$, $\mathcal{S}$, $\eps^t$, $I$, N, P, \texttt{False})\tcp*{Alg.~\ref{alg:projection}}
    }
  }

  $\eps^t \leftarrow$ \texttt{Project}($\mathcal{M}$, $\mathcal{S}$, $\eps^t$, $I$, N, P, \texttt{True})\tcp*{Alg.~\ref{alg:projection}}
  $\mathcal{S}.\texttt{reset}()$ \tcp*{Sec.~\ref{sec:progressive_proj}}
}
\end{algorithm}

%% file: algo/algo_proj.tex
\begin{algorithm}
\caption{Projection to Implementable Subspace via Inverse Design}
\label{alg:projection}

\KwIn{
Model $\mathcal{M}$; Sharpness scheduler $\mathcal{S}$; \\
Current metasurfaces $\epsilon$; Projection iteration count $I$; System level transfer matrix calibration $C$; number of meta atoms in each metasurface $N$; patch size $P$
}
\KwOut{Updated feasible design for all metasurfaces}

$\mathcal{S}_0 \leftarrow \texttt{copy}(\mathcal{S})$ \tcp*{Clone sharpness scheduler}
\For{$k \leftarrow 1$ \KwTo $K$}{
  $\mathcal{S} \leftarrow \mathcal{S}_0$ \tcp*{Reset sharpness scheduler}
  $\widehat{T_k} \leftarrow \mathcal{M}\texttt{.GetTM}(k)$ \tcp*{Target TM for layer $k$}
  \For{$j \leftarrow 1$ \KwTo $I$}{
    $\mathcal{S}.\texttt{step}()$ \tcp*{Sec.~\ref{sec:progressive_proj}}
    $s \leftarrow \mathcal{S}.\texttt{get}()$\;
    $T_k \leftarrow \texttt{ProbeTM}(\epsilon_k, s, P, N)$\tcp*{Sec.~\ref{sec:patch_probe_tm}}
    \texttt{DesignStep}($T_k$, $\widehat{T_k}$) \tcp*{Loss $\rightarrow$ Backward (adjoint simulation involved) $\rightarrow$ Step}
  }
}
\If(\tcp*[f]{Sec.~\ref{sec:fine_tune_entire}}){C is \texttt{True}}{
    \textbf{D} $\leftarrow$ $\mathcal{M}$\texttt{.GetNear2Far()}\;
    $\widehat{\mathcal{T}_{tot}} \leftarrow$ \textbf{D}\;
    \For(\tcp*[f]{Calc.target system TM}){$k \leftarrow 1$ \KwTo $K$}{
        $\widehat{\mathcal{T}_{k}} \leftarrow \mathcal{M}$\texttt{.GetTM}(k)\;
        $\widehat{\mathcal{T}_{tot}} \leftarrow \textbf{D} \, \widehat{\mathcal{T}_{k}} \, \widehat{\mathcal{T}_{tot}}$\;
    }
    $\mathcal{S} \leftarrow \mathcal{S}_0$ \tcp*{Reset sharpness scheduler}
    \For{$i \leftarrow 1$ \KwTo $I_{finetune}$}{
        $\mathcal{S}.\texttt{step}()$ \tcp*{Sec.~\ref{sec:progressive_proj}}
        $s \leftarrow \mathcal{S}.\texttt{get}()$\;
        $\mathcal{T}_{tot} \leftarrow$ \textbf{D}\;
        \For(\tcp*[f]{Calc.feasible system TM}){$k \leftarrow 1$ \KwTo $K$}{
            $\mathcal{T}_{k} \leftarrow \texttt{ProbeTM}(\epsilon_k, s, P, N)$ \tcp*{Sec.~\ref{sec:patch_probe_tm}}
            $\mathcal{T}_{tot} \leftarrow \textbf{D} \, \mathcal{T}_{k} \, \mathcal{T}_{tot}$\;
        }
        \texttt{DesignStep}($T_{tot}$, $\widehat{T_{tot}}$) \tcp*{Loss $\rightarrow$ Backward (adjoint simulation involved) $\rightarrow$ Step}
    }
}
\Return $\eps$\;
\end{algorithm}

%% file: algo/algo_tmProbe.tex
\begin{algorithm}
\caption{Patch-Based Transfer Matrix Sketching, see Sec.~\ref{sec:patch_probe_tm}}
\label{alg:patch_probe}

\KwIn{Metasurface permittivity $\epsilon$, binarization sharpness $s$, patch size $P$, number of meta-atoms $N$}
\KwOut{Transfer matrix $\widehat{T}$ of shape $M \times N$}

Initialize $\widehat{T} \leftarrow 0$ 	\tcp*{Empty transfer matrix}

\For{$i \leftarrow 0$ \KwTo $N - 1$}{
    $\epsilon_{\text{patch}} \leftarrow \texttt{ExtractPatch}(\epsilon, i, P, s)$ 	\tcp*{Patch centered at atom $i$}

    $r \leftarrow \texttt{NumericalSolver}(\epsilon_{\text{patch}}, \texttt{probe\_stimuli})$ 	\tcp*{Simulate near-field response}

    $[r_{\text{start}}, r_{\text{end}}] \leftarrow \texttt{ComputeRowIndices}(i, P, N)$ 	\tcp*{Determine response placement range}

    $\widehat{T}[r_{\text{start}} : r_{\text{end}}, i] \leftarrow r$ 	\tcp*{Insert patch response into transfer matrix}
}

\Return $\widehat{T}$

\end{algorithm}

%% file: figtex/fig_patchProbe.tex
\begin{figure}
    \centering
    \includegraphics[width=1\columnwidth]{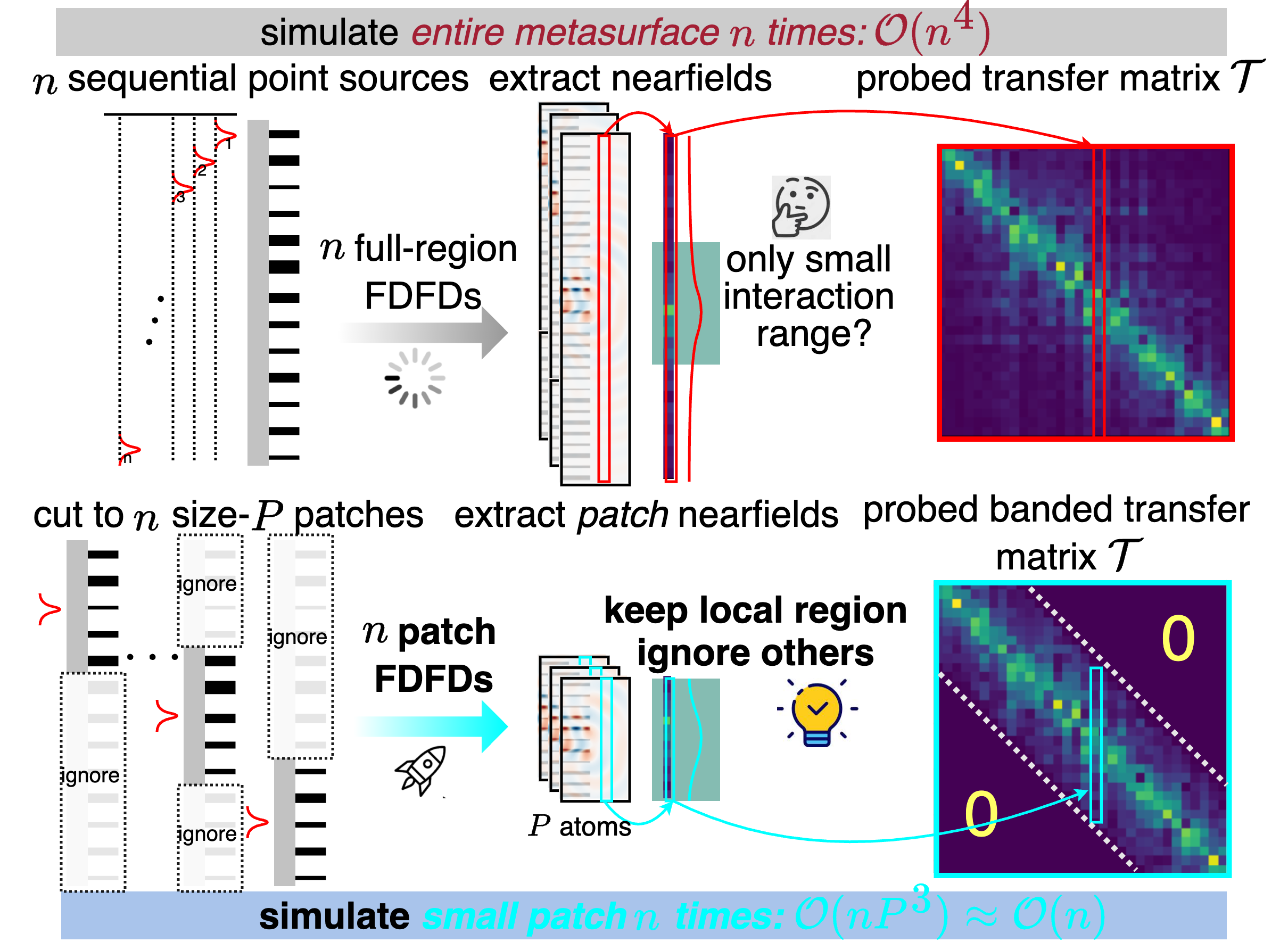}
    \caption{Proposed spatially-decoupled transfer matrix probing method cuts the metasurface into small patches for patch simulation that reduces complexity from cubic to linear.
    }
    \label{fig:patch_probe}
\end{figure}

%% file: figtex/fig_progressiveProj.tex
\begin{figure*}
    \centering
    \includegraphics[width=0.9\textwidth]{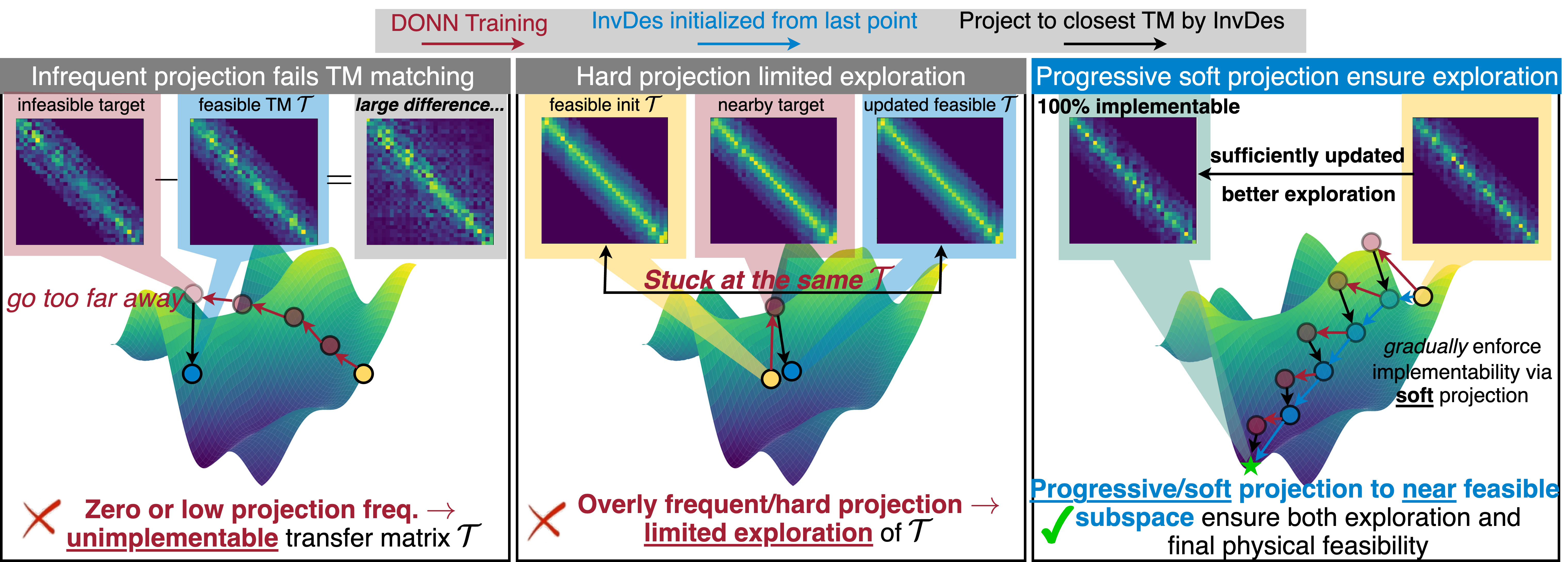}
    \caption{Our proposed \name framework enables both exploitation and physical feasibility.
   }
    \label{fig:progressive_proj}
    \vspace{-10pt}
\end{figure*}

%% file: doc/5_exp.tex
\vspace{-10pt}
\section{Evaluation}
\label{sec:evaluation}
\vspace{-5pt}
\subsection{Evaluation Settings}
\input{figtex/fig_expSettingErrorSrc}
\vspace{-10pt}
The hybrid DONN for classification tasks comprises an optical feature extractor followed by a lightweight digital classification head. 
The optical feature extractor operates analogously to a $3\times 3$ convolution. 
It has an inherent nonlinearity from the physical process of transmitting the optical field and measuring the resulting intensity. 
In our configuration, only TM polarization ($H_z$) is used without wavelength multiplexing, resulting in a single input channel. 
For the input data with multiple channels, such as the RGB images, we convert them to grayscale images. 
The output plane of DONN is partitioned into multiple regions, and the average light intensity within each region represents the outputs across different channels. 
In this work, we set the number of output channels to 4.

The optical feature extractor consists of nine inputs and 2 metasurfaces, where the pixels within the sliding window are phase-encoded ([0-1] maps to [0,$\pi$]). 
For simplicity, we adopt a one-dimensional (1D) metasurface configuration, where each metasurface consists of 32 Si meta-atoms with a period of $300\,\mathrm{nm}$ and a pillar height of $750\,\mathrm{nm}$. 
The pillar widths of the meta-atoms are trainable parameters, and the background material is air. 
The diffraction distances from the input waveguides to the first metasurface, between the two metasurfaces, and from the second metasurface to the receiver plane are all set to $4\,\mu\mathrm{m}$. 
The operating wavelength throughout the system is fixed at $850\,\mathrm{nm}$.

\vspace{-10pt}
\subsection{Performance Degradation Analysis}
\label{sec:degradation_analysis}
\vspace{-10pt}
Before presenting the detailed results, we analyze the sources of performance degradation due to transfer matrix relaxation and approximation. 
As illustrated in Fig.~\ref{fig:exp_setting_error_src}, several key factors contribute to this performance gap.

\noindent\ding{202}~\textbf{Projection Error}:~
If a physical realization of a metasurface could be found whose transfer matrix exactly matches the target transfer matrix obtained from unconstrained DONN training, we would expect to achieve identical performance on the training set. However, there is no guarantee that the target transfer matrices lie within the physically implementable subspace. Even with the proposed projection-based method, the implemented metasurface can only approximate the target transfer matrix, not replicate it exactly. 
This nonideal projection leads to performance degradation on the training set. 

\noindent\ding{203}~\textbf{TM Approximation Error}:~
The computational cost of projection is primarily determined by solving multiple linear systems of the form $Ax=b$, a process that scales proportionally with the number of stimuli injected to probe the transfer matrix. 
Given that the metasurface spans $9.6\,\mu\mathrm{m}$ and the simulation resolution is $50\,\mathrm{px}/\mu\mathrm{m}$, the full-resolution transfer matrices used during testing would be of size $480\times480$. 
Accurately probing such high-resolution transfer matrices would require injecting 480 point sources, which is prohibitively time-consuming during training. 
To mitigate this cost, we assume that the field value does not vary significantly within the region of a single meta-atom. 
Based on this assumption, we inject only one pulse source per meta-atom with the pulse width of a period, resulting in a $32\times$ reduction in simulation cost. 
Furthermore, we downsample the response, producing probed transfer matrices of size $480/S \times 32$, where $S$ denotes the downsampling factor. 
However, this downsampling introduces an additional \textbf{approximation error} compared to the true high-resolution transfer matrices, leading to further performance degradation on the training set.

\noindent\ding{204}~\textbf{Generalization Error}:~
NNs also have imperfect generalization that leads to a drop in test accuracy. 
However, we emphasize that the focus of this work is to \emph{train a DONN that achieves high performance under PDE constraints, not to improve its generalization}. 
Consequently, generalization error is \emph{out of the scope of this study}.

\vspace{-10pt}
\subsection{Patch Size Selection: Efficiency vs. Accuracy Trade-off}
\input{figtex/fig_patchSize}
\vspace{-10pt}
We first determine a critical hyperparameter: the patch size in patched TM probing, while for other hyperparameters, please see Appendix.~\ref{sec:other_hyperparam}
As discussed in Sec.~\ref{sec:patch_probe_tm}, our patched transfer matrix probing leverages the locality of near-field response to approximate that of the full system using overlapping patches. This approximation induces a band-diagonal structure in the transfer matrix, as shown in Fig.~\ref{fig:patchSize}.
Larger patch sizes improve the fidelity of transfer matrix probing by capturing more near-field interactions, but this comes at the cost of increased simulation time. 
To balance probing accuracy and efficiency, we set the patch size to 17 meta-atoms in all experiments.
It is also important to note that the optimal patch size depends on factors such as the wavelength, the meta-atom material, and the diffraction distance between the metasurface and the observation plane.

\vspace{-13pt}
\subsection{Main Result}
\input{tables/tab_mainResult}
\vspace{-10pt}
Table~\ref{tab:mainResult} compares the performance of DONNs trained using different methods on multiple benchmarks. 
Fashion-MNIST\cite{NN_FashionMNIST2017} and SVHN\cite{NN_svhn2011} represent image recognition (metric is \emph{accuracy and CE loss} which stands for the cross-entropy loss used to measure the distance between two distributions, the lower the closer), while Darcy Flow\cite{li2021fourier} shows the potential of DONNs for scientific machine learning, specifically for solving PDEs (metric is \emph{normalized L2-Norm}). 
For each method, we evaluate the test set performance using the final trained DONN model with real simulated responses of implemented metasurfaces.

We compare \name with three heuristic methods and end-to-end training. 
For the LPA-based method, only the phase mask is optimized, and the metasurfaces are designed using a precomputed look-up table. 
For \emph{Convolutional LPA}~\cite{wu2019neuromorphic}, the metasurface response is modeled as a convolution parameterized from meta-atom widths via interpolating the look-up table. 
The \emph{Smoothed Metasurface} method uses a regularization term to enforce similar meta-atom sizes across neighbors to mitigate errors caused by the breakdown of the local periodic approximation. 
Unlike the original method that directly regularizes the phases of adjacent meta-atoms, we adapt it by penalizing differences between adjacent columns of the transfer matrices. 
This encourages similar responses across neighboring meta-atoms under the same probe stimuli\cite{zhou2023physics}.

Because all heuristic methods oversimplify the complex metasurface responses, they exhibit significant performance degradation from training to testing, primarily due to \emph{projection optimality error}. 
In other words, the optimized transfer matrices produced by these methods are \emph{not physically realizable}. 
This leads to a large discrepancy between the target transfer matrices and their projections onto the implementable subspace. 
This mismatch ultimately causes a substantial performance drop during inference.
In contrast, the optimized transfer matrices produced by \name are guaranteed to lie within the implementable subspace.

\input{figtex/fig_inference_field}
To better illustrate how heuristic methods oversimplify light modulation, Fig.~\ref{fig:inference_field} compares the estimated and simulated optical fields within the diffractive feature extractor during inference. 
For systems with more and larger metasurfaces, please see Appendix~\ref{sec:vis_larger_sys}. 
\name accurately reproduces the field calculated by FDFD simulation, whereas LPA shows a significant field approximation error.

Consequently, the only sources of performance degradation of \name are the \ding{203} approximation error arising from the approximation of low-resolution to high-resolution transfer matrices, and the  \ding{204} generalization error from training to testing. 
Compared to \ding{202} projection error, these two factors have marginal impacts on performance degradation.

For the simulation-in-the-loop method\cite{khoram2019nanophotonic}, we estimate the training cost for a single epoch to be $\sim$178 hours for Fashion-MNIST with batch size 32, due to the large number of inverse designs where full-wave simulations are involved, making it prohibitively time-consuming. As a result, we consider this intractable and do not report results for this baseline.

\vspace{-10pt}
\subsection{Projection Configurations}
\vspace{-6pt}
\subsubsection{Progressive Soft Projection Schedule}
\vspace{-10pt}
As discussed in Sec.~\ref{sec:progressive_proj}, frequent projection steps that strictly enforce physical feasibility by mapping transfer matrices to a highly non-convex implementable subspace of binarized device patterns can severely hinder the optimization of the optical feature extractor. In such cases, the design often becomes trapped in a suboptimal local minimum or even remains close to the initialization.
As shown in Table~\ref{tab:progressive_sche}, when strict binarization is enforced at every projection step, the optical feature extractor stagnates with minimal updates. 

To address this, we adopt a relaxed binarization strategy that enables better exploration of the design space, leading to improved performance. 
Our proposed progressive soft projection gradually increases the binarization sharpness over time, allowing the metasurface to evolve meaningfully and contribute to higher classification accuracy, while still ensuring implementability through the final binarization.

While one could further apply relaxed binarization throughout the entire training process and enforce strict binarization only in the final epoch, this approach over-relaxes the trained matrices and makes the final performance highly sensitive to the optimization state at that specific moment. 
Also, as only the final design is guaranteed to be physically implementable, we lose the opportunity to select the best-performing design on the validation set before it is fully binarized.

\input{tables/tab_progressiveProj}

\vspace{-10pt}
\subsubsection{Projection Frequency}
\vspace{-10pt}
\input{tables/tab_projFreq}
Table~\ref{tab:proj_freq} compares different projection frequencies under different optimization runtime budgets. 
Projecting only once at the end of each training epoch results in significant performance degradation, as the unconstrained DONN training causes the target transfer matrices $\widehat{\mathcal{T}}$ to drift too far from the implementable subspace, resulting in poor alignment between their physically feasible projections and the targets. 
In contrast, increasing the projection frequency while proportionally reducing the number of iterations per projection maintains the overall runtime but significantly improves test accuracy.

Compared to projection frequency, the total optimization cost plays a relatively smaller role in maintaining high performance. 
As shown in Table~\ref{tab:proj_freq}, varying the total number of inverse design iterations per epoch, for example, halving or doubling it, has little impact on performance, considering that projections are already performed frequently. 
In our setting, we use 20 inverse design iterations per epoch with 10 projections per epoch, i.e., 2 adjoint gradient updates per projection.

\vspace{-13pt}
\subsubsection{Choose Appropriate Transfer Matrix Size}
\vspace{-10pt}
\input{figtex/fig_nearfield_ds_invdes_obj}
As mentioned previously in Sec.~\ref{sec:degradation_analysis}, there is a resolution gap between the transfer matrices $\widehat{\mathcal{T}}$ of size $480/S \times 32$, used to parameterize light modulation during DONN training, and the ground-truth transfer matrices of size $480 \times 480$. 
Because the optimization cost increases proportionally with the number of matrix columns, that is, using more fine-grained input sources to probe the transfer matrices, we do not increase the column count. 
Instead, we can adopt a finer resolution to represent the near-field response (matrix \textbf{rows}), thereby mitigating the approximation error. 
As shown in Fig.~\ref{fig:Downsample}, using a downsampling rate of 3 when we record the near-field can effectively minimize the performance drop. 
Note that varying the near-field downsampling rate does not affect the overall optimization cost.

\vspace{-10pt}
\subsubsection{Inverse Design Projection Objective}
\vspace{-10pt}
Figure~\ref{fig:Objective} shows how different projection objectives impact the performance drift from unconstrained target performance to the feasible performance. 
Layer-wise projection with our end-to-end matching effectively mitigates the performance drift caused by the inverse design imperfection.

\vspace{-13pt}
\subsection{Discussion}
\vspace{-6pt}
\input{figtex/fig_transferMetaline}

\subsubsection{Transfer Learning from Fashion-MNIST to SVHN}
\vspace{-10pt}
To demonstrate the effectiveness of the learned optical feature extractor, we performed transfer learning from Fashion-MNIST to SVHN. Despite the visual differences between the datasets, one consisting of grayscale images of clothing and the other of colorful house numbers, the transferred model shows faster convergence and improved performance compared to training from scratch, as shown in Fig.~\ref{fig:transfer}.

\vspace{-13pt}
\subsubsection{Generalization to Other Types of Meta-Optic Structures}
\vspace{-10pt}
Beyond metasurfaces, where light modulation is realized by varying the width of meta-atoms with fixed height, alternative approaches exist to control light behavior. 
For example, metalines achieve light modulation by fixing the width of meta-atoms and varying their height. 
Our proposed \name framework is compatible with such architectures and can be used to train DONNs implemented with metalines. 
This is demonstrated in Fig.~\ref{fig:metaline}, which shows the training dynamics of a metaline-based DONN.

%% file: figtex/fig_expSettingErrorSrc.tex
\begin{figure}
    \centering    \includegraphics[width=1\columnwidth]{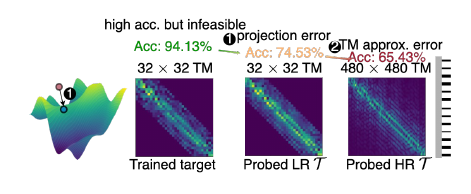}
    \vspace{-20pt}
    \caption{Error sources from \ding{202} inverse design projection error and \ding{203} transfer matrix approximation error causing the accuracy degradation from an ideally trained target matrix to a real response of the implemented metasurface.
    }
    \label{fig:exp_setting_error_src}
    \vspace{-10pt}
\end{figure}

%% file: figtex/fig_patchSize.tex
\begin{figure}[!]
    \centering
    \includegraphics[width=1\columnwidth]{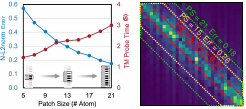}
    \vspace{-12pt}
    \caption{Different patch size trades off transfer matrix probing error and runtime.
    }
    \label{fig:patchSize}
    \vspace{-15pt}
\end{figure}

%% file: tables/tab_mainResult.tex
\begin{table}[]
\centering
\caption{Comparison across different DONN training methods on different benchmarks, our proposed \name consistently achieves best performance.
}
\resizebox{\columnwidth}{!}{%
\begin{tabular}{|c|ccccc|}
\hline
Benchmark                 & \multicolumn{1}{c|}{Baselines}                    & \multicolumn{1}{c|}{Train CE}                         & \multicolumn{1}{c|}{Train Acc}                       & \multicolumn{1}{c|}{Test CE}                      & Test Acc                        \\ \hline
                         & \multicolumn{1}{c|}{\cellcolor[HTML]{EFEFEF}\name} & \multicolumn{1}{c|}{\cellcolor[HTML]{EFEFEF}9.00E-02} & \multicolumn{1}{c|}{\cellcolor[HTML]{EFEFEF}96.85\%} & \multicolumn{1}{c|}{\cellcolor[HTML]{EFEFEF}0.45} & \cellcolor[HTML]{EFEFEF}88.44\% \\ \cline{2-6} 
                         & \multicolumn{1}{c|}{LPA\cite{luo2022metasurface}}                          & \multicolumn{1}{c|}{1.52E-01}                         & \multicolumn{1}{c|}{94.77\%}                         & \multicolumn{1}{c|}{2.25}                         & 58.79\%                         \\ \cline{2-6} 
                         & \multicolumn{1}{c|}{conv LPA~\cite{wu2019neuromorphic}}                     & \multicolumn{1}{c|}{1.69E-01}                         & \multicolumn{1}{c|}{94.11\%}                         & \multicolumn{1}{c|}{6.40}                         & 13.45\%                         \\ \cline{2-6} 
                         & \multicolumn{1}{c|}{smoothed metasurface~\cite{Li_lightridge_2024}}     & \multicolumn{1}{c|}{2.39E-02}                         & \multicolumn{1}{c|}{99.45\%}                         & \multicolumn{1}{c|}{22.61}                        & 16.21\%                         \\ \cline{2-6} 
\multirow{-5}{*}{Fashion-MNIST\cite{NN_FashionMNIST2017}} & \multicolumn{1}{c|}{sim-in-the-loop~\cite{khoram2019nanophotonic}}                      & \multicolumn{1}{c|}{Time out}                         & \multicolumn{1}{c|}{Time out}                        & \multicolumn{1}{c|}{Time out}                     & Time out                        \\ \hline
avg. improv.       & \multicolumn{5}{c|}{+66.67\%}                                                                                                                                                                                                                           \\ \hline
                         & \multicolumn{1}{c|}{\cellcolor[HTML]{EFEFEF}\name} & \multicolumn{1}{c|}{\cellcolor[HTML]{EFEFEF}3.16E-01} & \multicolumn{1}{c|}{\cellcolor[HTML]{EFEFEF}91.03\%} & \multicolumn{1}{c|}{\cellcolor[HTML]{EFEFEF}0.76} & \cellcolor[HTML]{EFEFEF}81.61\% \\ \cline{2-6} 
                         & \multicolumn{1}{c|}{LPA\cite{luo2022metasurface}}                          & \multicolumn{1}{c|}{2.99E-01}                         & \multicolumn{1}{c|}{91.70\%}                         & \multicolumn{1}{c|}{6.71}                         & 23.92\%                         \\ \cline{2-6} 
                         & \multicolumn{1}{c|}{conv LPA~\cite{wu2019neuromorphic}}                     & \multicolumn{1}{c|}{2.21E-01}                         & \multicolumn{1}{c|}{93.67\%}                         & \multicolumn{1}{c|}{2.51}                         & 11.51\%                         \\ \cline{2-6} 
                         & \multicolumn{1}{c|}{smoothed metasurface~\cite{Li_lightridge_2024}}   & \multicolumn{1}{c|}{5.98E-02}                         & \multicolumn{1}{c|}{98.55\%}                         & \multicolumn{1}{c|}{5.04}                         & 7.23\%                          \\ \cline{2-6} 
\multirow{-5}{*}{SVHN\cite{NN_svhn2011}}   & \multicolumn{1}{c|}{sim-in-the-loop~\cite{khoram2019nanophotonic}}                      & \multicolumn{1}{c|}{Time out}                         & \multicolumn{1}{c|}{Time out}                        & \multicolumn{1}{c|}{Time out}                     & Time out                        \\ \hline
avg. improv.       & \multicolumn{5}{c|}{+82.58\%}                                                                                                                                                                                                                           \\ \hline
                         & \multicolumn{1}{c|}{}                             & \multicolumn{2}{c|}{Train N-L2norm}                                                                          & \multicolumn{2}{c|}{Test N-L2norm}                                                  \\ \hline
                         & \multicolumn{1}{c|}{\cellcolor[HTML]{EFEFEF}\name} & \multicolumn{2}{c|}{\cellcolor[HTML]{EFEFEF}0.37}                                                            & \multicolumn{2}{c|}{\cellcolor[HTML]{EFEFEF}0.35}                                   \\ \cline{2-6} 
                         & \multicolumn{1}{c|}{LPA\cite{luo2022metasurface}}                          & \multicolumn{2}{c|}{0.32}                                                                                    & \multicolumn{2}{c|}{0.60}                                                           \\ \cline{2-6} 
                         & \multicolumn{1}{c|}{conv LPA~\cite{wu2019neuromorphic}}                     & \multicolumn{2}{c|}{0.33}                                                                                    & \multicolumn{2}{c|}{0.87}                                                           \\ \cline{2-6} 
                         & \multicolumn{1}{c|}{smoothed metasurface~\cite{Li_lightridge_2024}}   & \multicolumn{2}{c|}{0.33}                                                                                    & \multicolumn{2}{c|}{0.48}                                                           \\ \cline{2-6} 
\multirow{-5}{*}{Darcy Flow\cite{li2021fourier}}  & \multicolumn{1}{c|}{sim-in-the-loop~\cite{khoram2019nanophotonic}}                      & \multicolumn{2}{c|}{Time out}                                                                                & \multicolumn{2}{c|}{Time out}                                                       \\ \hline
avg. improv.       & \multicolumn{5}{c|}{+42.40\%}                                                                                                                                                                                                                           \\ \hline
total avg. improv.  & \multicolumn{5}{c|}{+63.88\%}                                                                                                                                                                                                                           \\ \hline
\end{tabular}
}
\label{tab:mainResult}
\end{table}

%% file: figtex/fig_inference_field.tex
\begin{figure}
    \centering
    \includegraphics[width=\columnwidth]{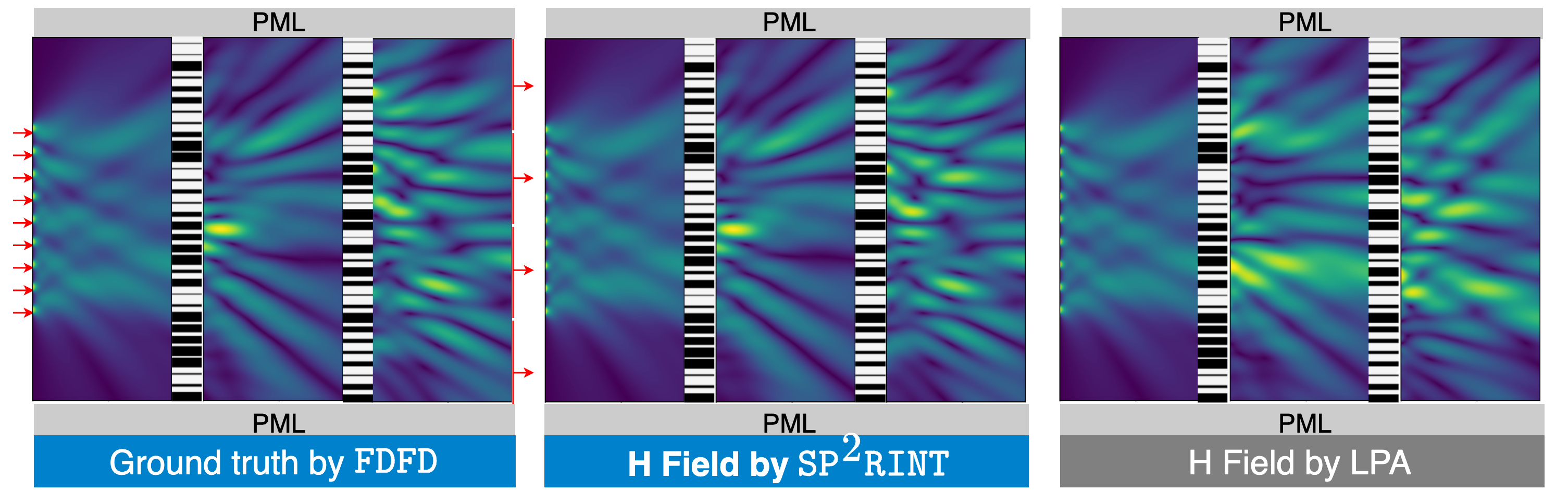}
    \caption{$|H_z|$ fields comparison on the 2-layer, 9-in-4-out metasurface system. 
    Our \name can accurately capture the metasurface transfer matrix, much more accurate than simple phase mask modeling using LPA, yielding almost the same H field amplitude calculated by FDFD.
   }
    \label{fig:inference_field}
    \vspace{-10pt}
\end{figure}

%% file: tables/tab_progressiveProj.tex
\begin{table}[]
\centering
\caption{
    Compare different projection sharpness schedules. 
    Progressively tightening the projection from a soft to a hard-binarization device ensures both exploration and physical feasibility. \emph{CE} is Cross-Entropy.
}
\resizebox{\columnwidth}{!}{%
\begin{tabular}{|c|c|c|c|}
\hline
\multirow{3}{*}{\begin{tabular}[c]{@{}c@{}}Sharpness\\ Schedule\end{tabular}}        
& \multirow{2}{*}{\begin{tabular}[c]{@{}c@{}}Per Projection\\
\includegraphics[width=0.27\columnwidth]{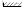}\end{tabular}}              
& \multirow{2}{*}{\begin{tabular}[c]{@{}c@{}}Per Epoch\\
\includegraphics[width=0.27\columnwidth]{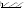}\end{tabular}}             
& \multirow{2}{*}{\begin{tabular}[c]{@{}c@{}}Per Train\\
\includegraphics[width=0.27\columnwidth]{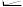}\end{tabular}}          \\
&&&\\
&&&\\
\hline
CE Loss $\downarrow$ & 0.579 & 0.428 & 0.954 \\ \hline
Accuracy $\uparrow$ & 86.19\% & 88.32\% & 80.82\% \\ \hline
\begin{tabular}[c]{@{}c@{}}Transfer\\ Matrix\end{tabular}
& \raisebox{-0.5\height}{\includegraphics[width=0.2\columnwidth]{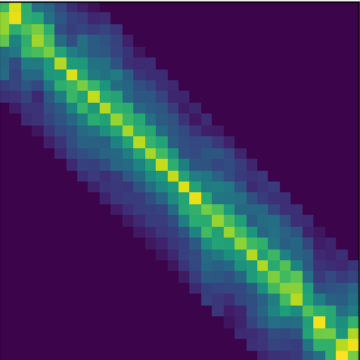}}
& \raisebox{-0.5\height}{\includegraphics[width=0.2\columnwidth]{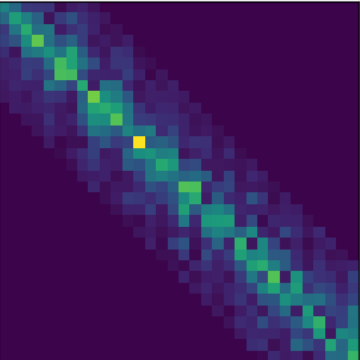}}
& \raisebox{-0.5\height}{\includegraphics[width=0.2\columnwidth]{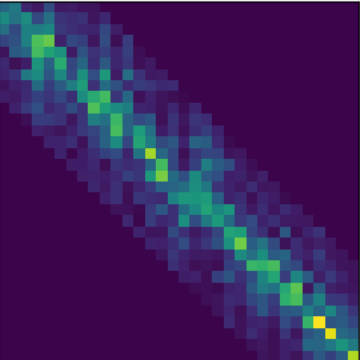}} \\ \hline
\end{tabular}
}
\label{tab:progressive_sche}
\end{table}

%% file: tables/tab_projFreq.tex
\begin{table}[]
\centering
\caption{Comparison between the projection schedule in one training epoch. 
By default, we define 1 unit budget as 20 iterations. 
Each iteration includes 2 times of forward transfer matrices probing, and corresponding adjoint simulation.
}
\resizebox{0.9\columnwidth}{!}{%
\begin{tabular}{|c|c|c|c|c|}
\hline
Cost                        & Proj. frequency                               & Proj. iters                          & CE $\downarrow$                              & Accuracy $\uparrow$                        \\ \hline
                            & 1 / epoch                                   & 20                                 & 1.190                         & 56.04\%                         \\ \cline{2-5} 
                            & 2 / epoch                                   & 10                                 & 0.939                         & 71.68\%                         \\ \cline{2-5} 
                            & 4 / epoch                                   & 5                                  & 0.360                         & 87.59\%                         \\ \cline{2-5} 
                            & 5 / epoch                                   & 4                                  & 0.444                         & 84.99\%                         \\ \cline{2-5} 
                            & \cellcolor[HTML]{EFEFEF}\textbf{10 / epoch} & \cellcolor[HTML]{EFEFEF}\textbf{2} & \cellcolor[HTML]{EFEFEF}0.352 & \cellcolor[HTML]{EFEFEF}89.54\% \\ \cline{2-5} 
\multirow{-6}{*}{budget $\times$1} & 20 / epoch                                  & 1                                  & 0.354                         & 88.64\%                         \\ \hline
budget $\times$1/2                 & 5 / epoch                                   & 2                                  & 0.514                         & 87.64\%                         \\ \hline
budget $\times$1/2                 & 10 / epoch                                  & 1                                  & 0.406                         & 88.76\%                         \\ \hline
budget $\times$2                   & 10 / epoch                                  & 4                                  & 0.452                         & 83.91\%                         \\ \hline
\end{tabular}
}
\label{tab:proj_freq}
\end{table}

%% file: figtex/fig_nearfield_ds_invdes_obj.tex
\begin{figure}
    \centering
    \vspace{-10pt}
    \subfloat[]{\includegraphics[width=0.54\columnwidth]{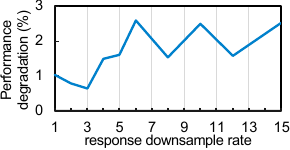}
    \label{fig:Downsample}}
    \subfloat[]{\includegraphics[width=0.44\columnwidth]{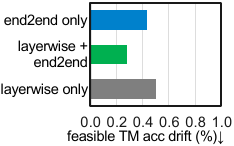}
    \label{fig:Objective}}
    \vspace{-10pt}
    \caption{(a) Lower near-field downsampling rate leads to smaller performance drop.
    (b) Combination of layerwise and end-to-end inverse design gives the best projection accuracy.
    }
    \label{fig:nearfield_ds_invdes_obj}
    \vspace{-10pt}
\end{figure}

%% file: figtex/fig_transferMetaline.tex
\begin{figure}
    \centering
    \subfloat[]{\includegraphics[width=0.49\columnwidth]{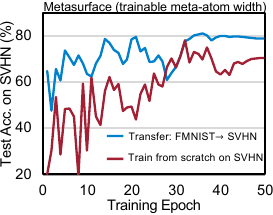}
    \label{fig:transfer}
    }
    \subfloat[]{\includegraphics[width=0.49\columnwidth]{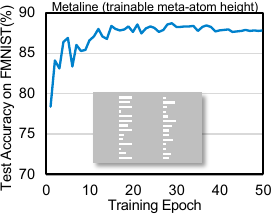}
    \label{fig:metaline}
    }
    \vspace{-10pt}
    \caption{(a) Transfer learning of DONNs from Fashion-MNIST to SVHN shows faster convergence and higher performance than training on SVHN from scratch.
    (b) \name can be generally applied to other devices such as metalines with trainable meta-atom height.
    }
    \label{fig:transfer_metaline}
    \vspace{-10pt}
\end{figure}

%% file: doc/6_conclu.tex
\vspace{-10pt}
\section{Conclusion}
\vspace{-10pt}
We propose \name, a scalable and physically grounded framework for training DONN. 
\name combines PDE-constrained optimization with spatially decoupled transfer matrix probing and progressive projection to ensure physical realizability without requiring per-iteration full-wave simulations.
Evaluation shows that \name achieves average test accuracy improvements of 63.88\% over widely used heuristic training methods and delivers $1825\times$ speed-up compared to the simulation-in-the-loop training.
These results highlight \name's potential to bridge the gap between high-performance deep learning and physically implementable nanophotonic hardware, enabling scalable, generalizable, and deployable optical neural systems.
\vspace{-15pt}
\section{Acknowledgment}
\vspace{-15pt}
The authors would like to acknowledge the NVIDIA Academic Grant Program.

\vspace{-15pt}
\section{Data Availability}
\vspace{-10pt}
The data that support the findings of this study are openly available in \href{https://github.com/ScopeX-ASU/SP2RINT}{\name}

%% file: doc/7_appendix.tex
\section{Appendix}
\subsection{Determine Other Hyperparameters}
\label{sec:other_hyperparam}
Besides the most important hyperparameter, the patch size, here we show how we choose other hyperparameters, i.e., the convolution kernel size that determines the number of input pixels to DONNs, the output channel number that determines the number of detectors on DONN output plane, the input port spacings (the spacing between adjacent pixels that are fed into DONNs), and the input pixel width.
\input{figtex/fig_conv_ks_output_channel}
\input{figtex/fig_wg_config}

\noindent{\textbf{Convolution Kernel Size}}.~
We sweep three convolution kernel sizes, and we chose a kernel size of 3, i.e., each time 3$\times$3=9 pixels are fed into a DONN with 9 input ports per sliding window, which delivers the best cross-entropy (CE) loss and test accuracy as shown in Fig.~\ref{fig:conv_kernel_size}. 
We do not consider a larger kernel size since a larger kernel size, e.g., kernel size of 5, requires at least $5^2 = 25$ input ports, which is too many to place considering that the metasurface only has a diameter of $32 \times 0.3 = 9.6~\mu m$.

\noindent{\textbf{Output Channel Number}}.~
The output channel count equals the output detector count in DONNs.
There is a trade-off between expressivity and robustness in the choice of output channel count.
A larger number of output channels represents higher expressivity while compromising the robustness to the errors mentioned in Sec.~\ref{sec:degradation_analysis}.
Recall that we average the light intensity within one detector region as the output.
Too many ports mean each port covers a small region that is susceptible to non-ideal fluctuations.
Furthermore, if output ports are too crowded/close, their transfer functions are highly correlated, which does not necessarily increase the effective channel capacity.
We sweep the number of output channels and find that the output channel number of 4 delivers the best CE loss and test accuracy, as shown in Fig.~\ref{fig:output_channel_num}.

\noindent{\textbf{Input Port Spacing}}.~
The spacing between the input ports is also a critical hyperparameter. 
Here we sweep different intervals from 0.4 $\mu m$ to 0.8 $\mu m$ and we observe that $0.4\mu m$ provides the best CE loss and test accuracy as shown in Fig.~\ref{fig:input_wg_interval}.

\noindent{\textbf{Input Port Width}}.~
The input port width determines the light pulse width that encodes each pixel.
We sweep the input port width, ranging from 0.1 $\mu m$ to 0.5 $\mu m$, and it shows that 0.2 $\mu m$ provides the best CE loss and test accuracy, as shown in Fig.~\ref{fig:input_wg_width}.

\vspace{-10pt}
\subsection{Evaluate \name on Larger DONN Systems}
\label{sec:vis_larger_sys}
\vspace{-10pt}
\input{figtex/fig_runtime_compare}
\input{figtex/fig_error_accumulation}
\input{figtex/fig_larger_inference_field_32}
\input{figtex/fig_larger_inference_field_64}
\input{figtex/fig_larger_inference_field_128}

\input{figtex/fig_larger_inference_field_160}

Note that for the results shown here, we use a patch size of 53 meta-atoms for metasurfaces with 64, 128, and 160 meta-atoms, and a patch size of 27 for the 32-meta-atom case, instead of the default size of 17 used in Sec.~\ref{sec:evaluation}. 

Figure~\ref{fig:run_time_compare} compares the transfer matrix probing time between full-size simulations and the proposed \name across different metasurface sizes. 
With patched transfer matrix probing, \name~achieves linear probing computational complexity, in contrast to the polynomial complexity of full-size simulation, making it scalable to larger systems.
All patches can be potentially simulated in parallel, which theoretically gives a constant simulation runtime, the same as simulating a single patch.

Figure~\ref{fig:error_accumulation} shows the normalized $L_2$ error between the input field magnitude at each metasurface layer and the ground truth in 6-layer cascaded systems of varying sizes. While both LPA and \name~exhibit error accumulation as light propagates deeper into the system, \name~consistently achieves significantly more accurate field estimation than LPA.

Fig.~\ref{fig:larger_inf_field_32}, Fig.~\ref{fig:larger_inf_field_64}, Fig.~\ref{fig:larger_inf_field_128}, and Fig.~\ref{fig:larger_inf_field_160} visualize the light field magnitude $|H_z|$ in 6-layer diffractive systems composed of metasurfaces with 32, 64, 128, and 160 meta-atoms, respectively.
Slices of intermediate fields are extracted and plotted for better comparison.
In all cases, \name consistently produces more accurate field estimation than LPA.

%% file: figtex/fig_conv_ks_output_channel.tex
\begin{figure}
    \centering
    \vspace{-10pt}
    \subfloat[]{\includegraphics[width=0.5\columnwidth]{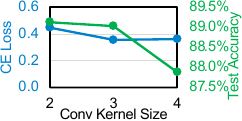}
    \label{fig:conv_kernel_size}}
    \subfloat[]{\includegraphics[width=0.5\columnwidth]{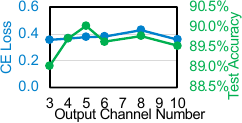}
    \label{fig:output_channel_num}}
    \vspace{-10pt}
    \caption{(a) Choose a convolution kernel size of 3 to deliver the best CE loss and test accuracy.
    (b) Choose the output channel number as 4, which delivers the best CE loss and test accuracy.
    }
    \label{fig:conv_ks_out_channel_num}
    \vspace{-10pt}
\end{figure}

%% file: figtex/fig_wg_config.tex
\begin{figure}
    \centering
    \vspace{-10pt}
    \subfloat[]{\includegraphics[width=0.5\columnwidth]{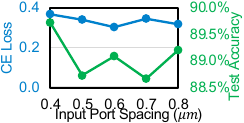}
    \label{fig:input_wg_interval}}
    \subfloat[]{\includegraphics[width=0.5\columnwidth]{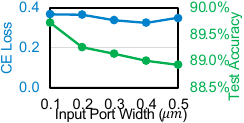}
    \label{fig:input_wg_width}}
    \vspace{-10pt}
    \caption{(a) Input port spacing of 0.4 $\mu m$ delivers best CE loss and test accuracy.
    (b) Input port width of 0.2 $\mu m$ delivers best CE loss and test accuracy.
    }
    \label{fig:input_wg_config}
    \vspace{-10pt}
\end{figure}

%% file: figtex/fig_runtime_compare.tex
\begin{figure}
    \vspace{15pt}
    \centering    
    \includegraphics[width=0.65\columnwidth]{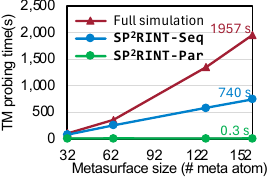}
    \caption{\name~achieves linear transfer matrix $\mathcal{T}$ probing time with respect to metasurface size, in contrast to the polynomial complexity of full-size simulation. With sufficient computational resources for patch-level parallelism, \name~can further reduce probing time toward constant complexity.
}
    \label{fig:run_time_compare}
\end{figure}

%% file: figtex/fig_error_accumulation.tex
\begin{figure*}
    \centering
    \includegraphics[width=1\textwidth]{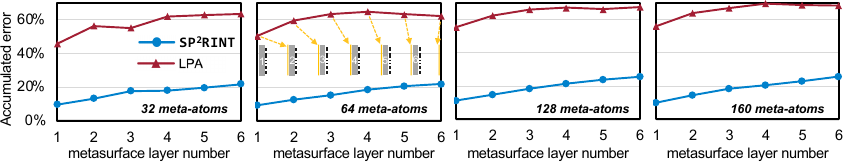}
    \caption{Normalized $L_2$ error between magnitude of the input field at each metasurface layer and the ground truth, measured across 6-layer cascaded systems of varying metasurface sizes. While both SP$^2$RINT and LPA experience error accumulation, SP$^2$RINT consistently achieves significantly lower error by honoring the underlying PDE constraints during training.
}
    \label{fig:error_accumulation}
    \vspace{-10pt}
\end{figure*}

%% file: figtex/fig_larger_inference_field_32.tex
\begin{figure*}
    \centering
    \subfloat[]{\includegraphics[width=1\textwidth]{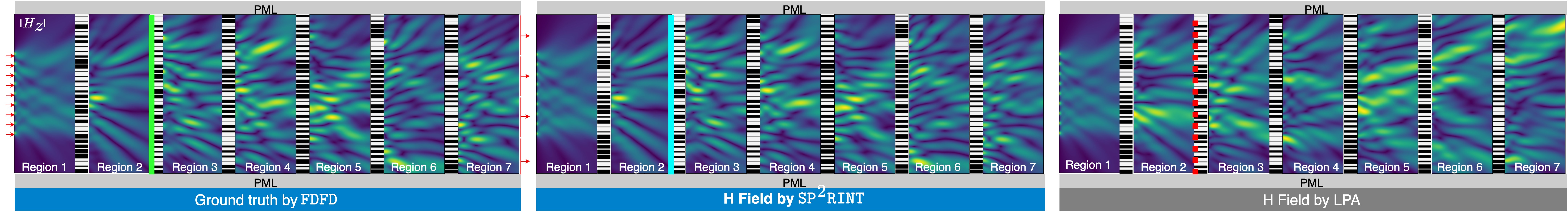}
    \label{fig:larger_inf_field_32a}}\\
    \vspace{-4pt}
    \subfloat[]{\includegraphics[width=1\textwidth]{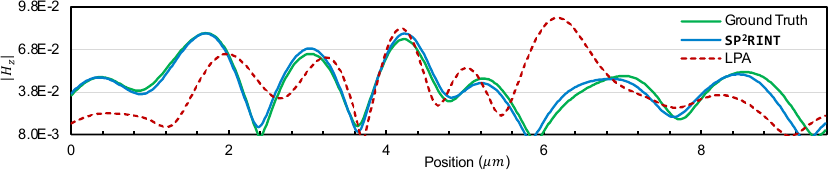}
    \label{fig:larger_inf_field_32b}}
    \caption{(a) Overall magnetic field $H_z$, and
    (b) Magnetic field $H_z$ slice comparison between ground truth given by FDFD, \name, and LPA for a 6-layer diffraction system consisting of 32-meta-atoms metasurfaces}
    \label{fig:larger_inf_field_32}
    \vspace{-10pt}
\end{figure*}

%% file: figtex/fig_larger_inference_field_64.tex
\begin{figure*}
    \centering
    \subfloat[]{\includegraphics[width=1\textwidth]{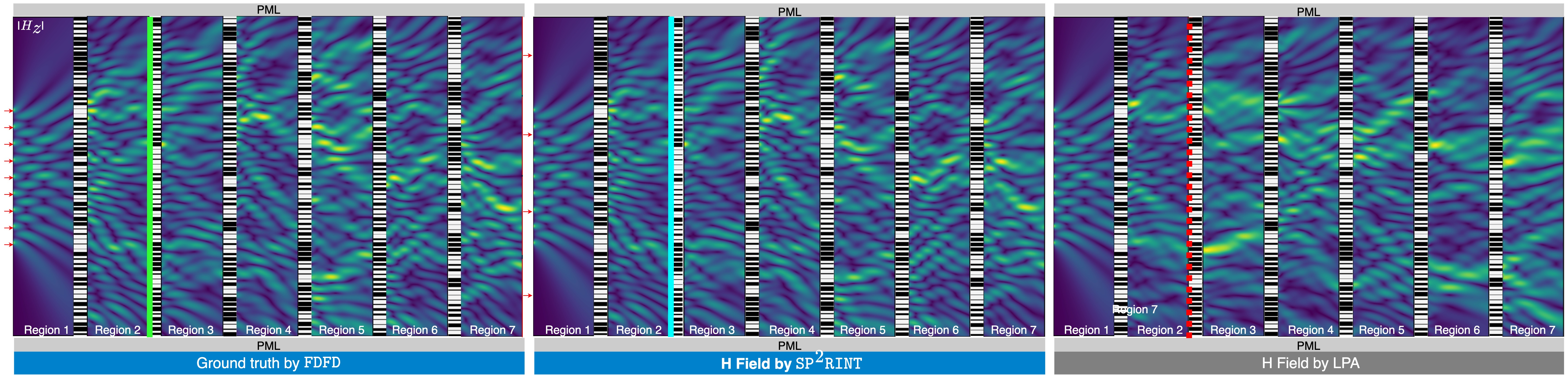}
    \label{fig:larger_inf_field_64a}}\\
    \vspace{-4pt}
    \subfloat[]{\includegraphics[width=1\textwidth]{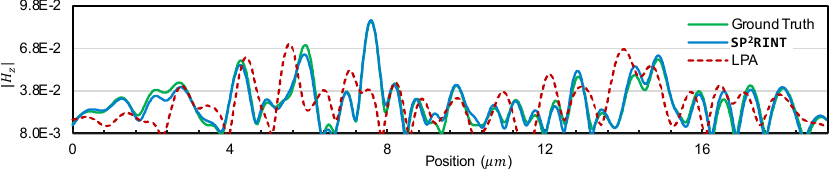}
    \label{fig:larger_inf_field_64b}}
    \caption{(a) Overall magnetic field $H_z$, and
    (b) Magnetic field $H_z$ slice comparison between ground truth given by FDFD, \name, and LPA a 6-layer diffraction system consisting of 64-meta-atoms metasurfaces}
    \label{fig:larger_inf_field_64}
    \vspace{-10pt}
\end{figure*}

%% file: figtex/fig_larger_inference_field_128.tex
\begin{figure*}
    \centering
    \subfloat[]{\includegraphics[width=0.8\textwidth]{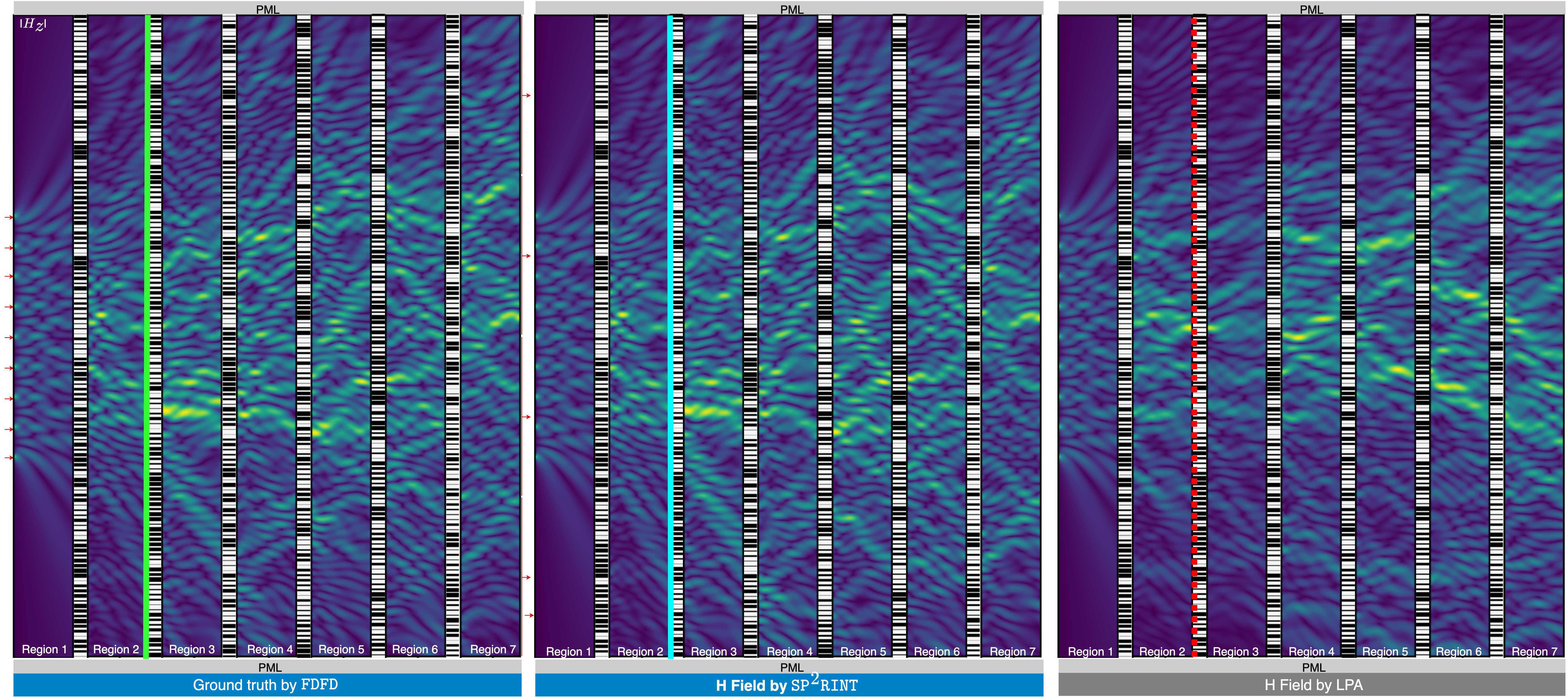}
    \label{fig:larger_inf_field_128a}}\\
    \vspace{-4pt}
    \subfloat[]{\includegraphics[width=0.8\textwidth]{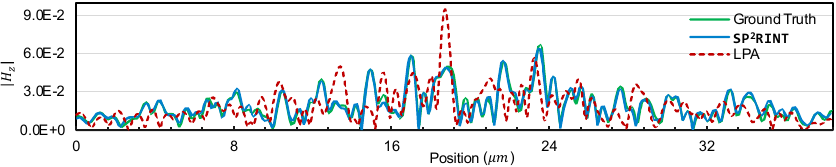}
    \label{fig:larger_inf_field_128b}}
    \vspace{-10pt}
    \caption{(a) Overall magnetic field $H_z$, and
    (b) Magnetic field $H_z$ slice comparison between ground truth given by FDFD, \name, and LPA a 6-layer diffraction system consisting of 128-meta-atoms metasurfaces}
    \label{fig:larger_inf_field_128}
    \vspace{-10pt}
\end{figure*}

%% file: figtex/fig_larger_inference_field_160.tex
\begin{figure*}
    \centering
    \subfloat[]{\includegraphics[width=0.8\textwidth]{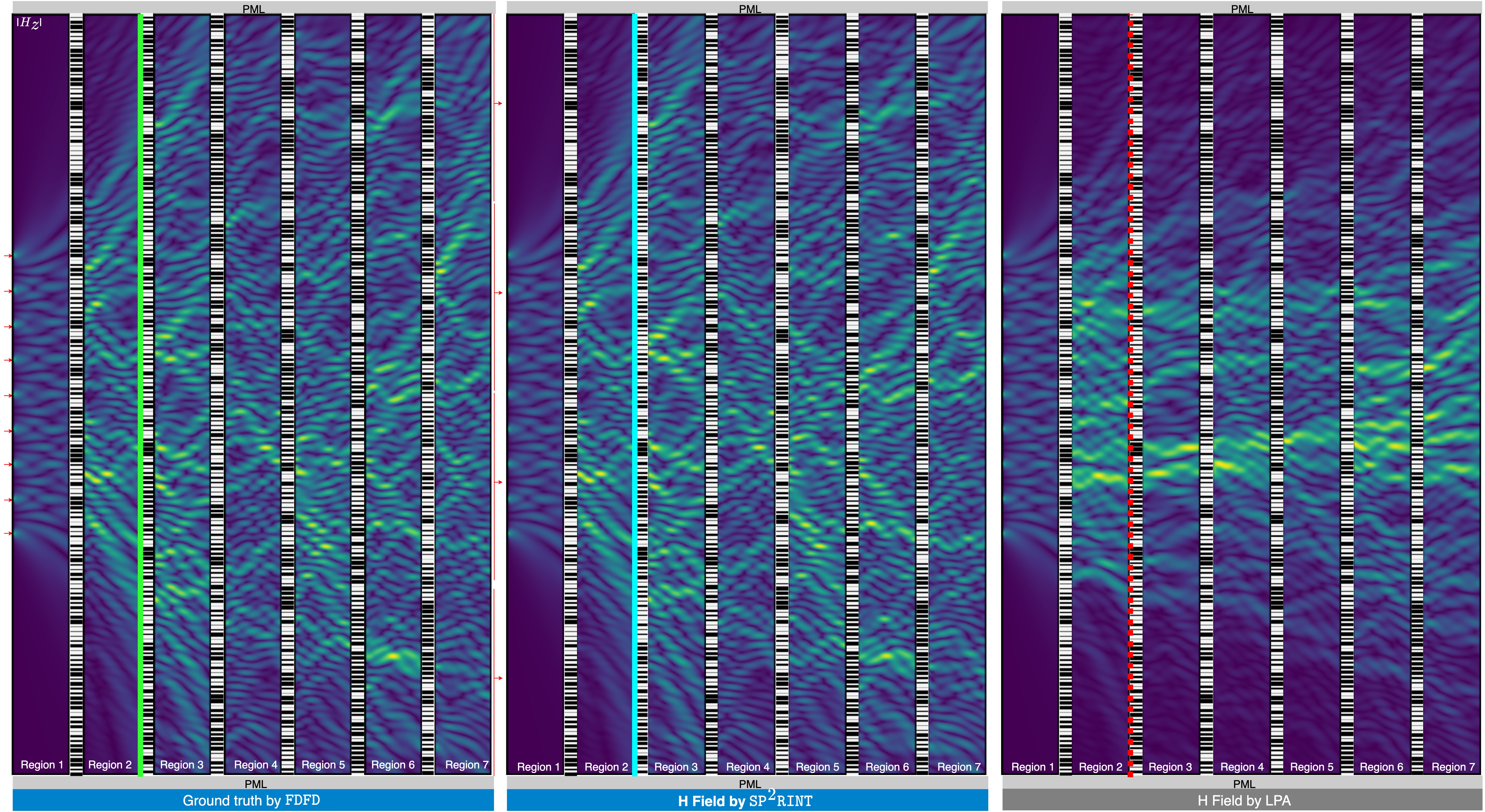}
    \label{fig:larger_inf_field_160a}}\\
    \vspace{-4pt}
    \subfloat[]{\includegraphics[width=0.8\textwidth]{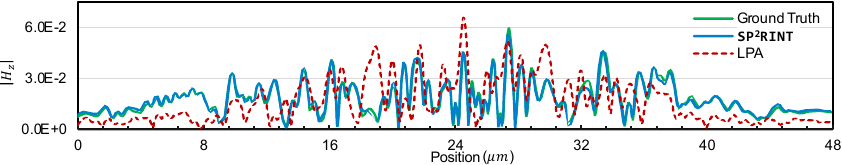}
    \label{fig:larger_inf_field_160b}}
    \vspace{-10pt}
    \caption{(a) Overall magnetic field $H_z$, and
    (b) Magnetic field $H_z$ slice comparison between ground truth given by FDFD, \name, and LPA a 6-layer diffraction system consisting of 160-meta-atoms metasurfaces}
    \label{fig:larger_inf_field_160}
    \vspace{-10pt}
\end{figure*}